\documentclass[
english,twoside,aps,prl,superscriptaddress,nofootinbib,superscriptaddress,nobibnotes]{revtex4-2}
\usepackage[T1]{fontenc}
\usepackage{amsmath}
\usepackage{amssymb}
\usepackage{graphicx}
\usepackage[utf8]{inputenc}
\usepackage{array}
\usepackage{parallel}
\usepackage{tikz,pgfplots}
\usepackage{subcaption}
\usepackage{hyperref}
\usepackage[normalem]{ulem}
\usepackage{setspace}

\pdfpageheight\paperheight
\pdfpagewidth\paperwidth

\usepackage{babel}
\definecolor{byzantine}{rgb}{0.74, 0.2, 0.64}\definecolor{vividviolet}{rgb}{0.62, 0.0, 1.0}
\definecolor{LightBlue}{rgb}{0.667, 0.847, 1}

\newcommand{\Jav}[1]{{\color{black}{#1}}}

\newcommand{\Mout}[1]{}
\providecommand{\tabularnewline}{\\}

\pgfplotsset{compat=1.16} 
\begin{document}

\title{Emergence of Dark Energy 
from topology and chiral spinors}

\author{J. Lorca Espiro}%
\email{javier.lorca@ufrontera.cl}
\affiliation{Departamento de Ciencias F\'{i}sicas, Facultad de Ingenier\'{i}a y Ciencias, Universidad de La Frontera, Avda. Francisco Salazar 01145, Casilla 54-D Temuco, Chile.}%
\author{Yerko V\'{a}squez}
\email{yvasquez@userena.cl}
\affiliation{Departamento de F\'{i}sica, Facultad de Ciencias, Universidad de La Serena, Avenida Cisternas 1200, La Serena, Chile.}
\author{M. Le Delliou}
\email{delliou@lzu.edu.cn,Morgan.LeDelliou.IFT@gmail.com}
\affiliation{-Institute of Theoretical Physics \& Research Center of Gravitation, Lanzhou University, Lanzhou 730000, China\\
	-Key Laboratory of Quantum Theory and Applications of MoE, Lanzhou University, Lanzhou 730000, China\\
	-Lanzhou Center for Theoretical Physics \& Key Laboratory of Theoretical Physics of Gansu Province, Lanzhou University, Lanzhou 730000, China} 	
	\affiliation{Instituto de Astrof\'isica e Ci\^encias do Espa\c co, Universidade de Lisboa,
	Faculdade de Ci\^encias, Ed.~C8, Campo Grande, 1769-016 Lisboa, Portugal}
 \affiliation{Universit\'e de Paris-Cit\'e, APC-Astroparticule et Cosmologie (UMR-CNRS 7164), 
 F-75006 Paris, France.}


\begin{abstract}
Under the existence of a massless spinor with respect to the total connection in a spacetime modeled as a Lorentzian manifold with internal boundaries, such as finite volume semi-classical Black Holes, we show that a topological mechanism naturally induces terms in the Einstein-Cartan gravitational action that can be interpreted as General Relativity with dark energy\Jav{\Mout{and some dark fluid}}. This may alleviate the problems of dark energy. \Jav{\Mout{The dark fluid term remains to be further analyzed.}} The topological information is carried by a harmonic 1-form associated to the first co-holomology group of the spacetime, which induces a spacetime contortion acting on the horizontal bundle.
\end{abstract}

\flushbottom
\maketitle
\tableofcontents
%
%


\section{Introduction.}



In Einstein's theory of gravitation, spacetime is mathematically described as a topological space that is also a differentiable manifold. Spacetime curvature is a central concept, illustrating how spacetime bends in response to mass and energy. The framework of the present paper allows spacetime to exhibit both curvature and torsion, depending on the connections defined on the manifold. While curvature describes the bending of spacetime, torsion pertains to its twisting or rotational aspects.

The study of theories of gravity with non null spacetime torsion has a very long history. The Einstein-Cartan-Sciama-Kibble (ECSK) theory of gravity \cite{cartan:1922}, first introduced by Cartan in 1922, is a modification of General Relativity (GR) that allows 
spacetime to have torsion in addition to curvature, while GR describes the gravitational interaction only in terms of spacetime curvature. The theory was rediscovered by Sciama and Kibble independently in the 1960s \cite{sciama, Kibble:1961ba}. In the ECSK theory, torsion is related to the density of intrinsic angular momentum \cite{Trautman:2006fp} and it is supposed to represent additional degrees of freedom of the gravitational field; therefore, there might be new physics associated to 
spacetime torsion. However, torsion vanishes outside matter, and does not propagate in vacuum. At the macroscopic level, where spins vanish, it coincides with GR, while at microscopic level it shows different results. Torsion also naturally appears when conformal rescalings of spinors are considered, through a complex conformal factor \cite{Penrose:1983mf}. For an extensive review on various aspects of classical theory of gravity with torsion refer to \cite{Hehl}, while the quantum aspects of the torsion are discussed in \cite{Shapiro:2001rz}.

Generalizations of the ECSK action have been considered that allow propagating torsion \cite{Neville:1979rb}. In such theories one could, in principle, have long-range torsion mediated interactions. For instance, one can consider higher-order curvature corrections or couple additional fields to the curvature, resulting in torsion being a propagating field rather than one which vanishes outside matter. Among theories of gravity with propagating torsional degrees of freedom we can mention the Poincar\'e gauge theory of gravity \cite{Blagojevic:2002du, Blagojevic:2013xpa}, which have both curvature and torsion, and the teleparallel equivalent to GR \cite{unzicker2005}, an equivalent formulation of GR proposed by Einstein in 1928, where the torsion represents the gravitational field while the curvature vanishes; for a comprehensive review see \cite{Aldrovandi:2013wha}. In addition, scalar fields coupled to curvature can serve as generators of torsion. In particular the coupling of scalar fields with characteristic classes have received great attention, because this last type of interactions manifests in various scenarios, such as the dimensional reduction observed in the low-energy limits of string theory and of loop quantum gravity. Specifically, in theories like Einstein-Gauss-Bonnet gravity \cite{Kanti:1995vq}, a non-minimal coupling arises between the dilaton and the topological terms associated with the Euler class, and in Chern-Simons gravity where a scalar field is coupled to the Pontryagin density, which is an effective extension of GR motivated by anomaly cancellation in particle physics and string theory; for a review of Chern-Simons gravity see \cite{Alexander:2009tp}. In Refs. \cite{Toloza:2012sn, Espiro:2014uda, Espiro:2019xkd,LeDelliou:2019mus} the authors have explored the repercussions of incorporating a term into the action that constitutes the product of a scalar field with the Euler classes, the Pontryagin classes and the Chern type Nieh-Yan classes \cite{Nieh:1981ww, Nieh:2007zz} in the presence of torsion. The last gives rises to a topological invariant immediately null in the absence of torsion, that have great importance in different branches of physics such as GR \cite{Date:2008rb, Chandia:1997hu, Mardones:1990qc, Sengupta:2013lxa, Bombacigno:2019nua} and condensed matter physics \cite{Nissinen:2019kld}.

Despite the success of GR, the occurrence of singularities and the limitations coming from the low-energy regime suggest that it is imperative to explore theories of gravity beyond GR. In this regard, theories with torsional degrees of freedom have been considered in the study of compact objects \cite{Obukhov:2019fti, Cembranos:2016gdt, Cvetkovic:2022qpt, Boos:2023xoq} and to solve cosmological problems of the very early or present universe \cite{ Magueijo:2012ug, MINKEVICH1980232, delaCruzDombriz:2021nrg, Poplawski:2012ab, Bolejko:2020nbw, Poplawski:2011jz, Akhshabi:2017lyg}. For example, in \cite{Cembranos:2016gdt} a static spherically symmetric solution was found. The solution describes a modified Schwarzschild metric, where torsion provides an extra term in the metric. In Ref.~\cite{Boos:2023xoq} regular black holes (hereafter BHs) were constructed from a confined spin connection in the Poincar\'e gauge theory of gravity. On the other hand, in \cite{Bolejko:2020nbw} the authors showed that, when torsion is present, the cosmic duality relation between the angular diameter distance and the luminosity distance is broken. Models with torsion that can replace the big bang singularity with a cusp-like bounce at finite minimum scale factor were proposed in \cite{Poplawski:2011jz}, while the effects of spin and torsion can also lead to an inflationary phase without the need of additional fields \cite{Akhshabi:2017lyg}. Furthermore, torsional degrees of freedom have been proposed as alternative to dark energy \cite{Espiro:2019xkd,LeDelliou:2019mus, Kranas:2018jdc, Benisty:2021sul}. For a chronological review of the literature on non-Riemannian cosmological models, see \cite{Puetzfeld:2004yg}.


In this paper we propose a mechanism of topological origin which induces naturally terms in the gravitational action, that can be interpreted as dark energy \Jav{\Mout{and a dark fluid,}} under the considerations of a spacetime with internal boundaries and the inclusion of spinor degrees of freedom in the framework. The essential topological features of the construction is measured by the $\pi_3 (\mathcal{M})$ homotopy group, where the relevant topological information is contained in the equivalence class of harmonic 1-forms associated to the co-homology group $H^1(\mathcal{M})$ when strict Neumann boundary conditions are considered. We show that this 1-form naturally induces a spacetime contortion 1-form, and a torsion 2-form, which by the parallel spinor hypothesis acts solely on the horizontal bundle, that allows 
topological information to enter the dynamics of the gravitational theory.

The manuscript is organized as follows: in Sec.~\ref{sec:structure} we introduce the action of the model we study in this paper. From it, we obtain its field equations. The rest of the sections study the field equations in more detail. In Sec.~\ref{sec:scalarfield} we present how geometrical and topological information can emerge directly from the scalar field equations. In ~\ref{sec:spinconnection} we link the spin connection field equation and its torsion part with the topological information. In Sec ~\ref{sec:parallelspinors} we focus on the spinor field equation; the total conservation of spinors, the \textit{parallel spinor hypothesis}, leads to a Higgs-like emergence of spinor effective mass. In Sec. ~\ref{sec:curvature}, the frame field equations, including all above effects, yield a behaviour formally equivalent to GR with a dynamical \textit{dark energy} \Jav{\Mout{and a \textit{dark fluid} associated to a scalar field}}. We end with Sec.~\ref{sec:final}, where comments and final remarks are presented.

\section{\label{sec:structure} Structure and Generalities.}

In this section, we set the stage of the model we propose, describing the spacetime topology and other mathematical structures that are needed for the model. We start from a spacetime with curvature, torsion and inner holes, and workout the consequences on the model's action and field equations.

\subsection{\label{subsec:princbundle} Principal bundle and boundary of spacetime.}

From general to particular, we begin by assuming the existence of $\pi: P \rightarrow \mathcal{M}$, that is of a principal $G$-bundle structure over some oriented compact Lorentzian $4$-manifold $\mathcal{M}$. In this setting, a principal $G$-connection will be a differential $1$-form with values in the Lie algebra $\mathfrak{g}$ of $G$. If we also require this connection to be equivariant in the principal Lie group action, we can consider this connection to be represented by an Ehresmann connection.

We are then interested in $\mathcal{M}$ to be compact, which we take to coincide with the \textit{observable physical universe}, defined by: 
\begin{align}\label{regiondef}
{\mathcal{M}} \cong \mathcal{N} \setminus \mathcal{X}  \quad  \text{where} \quad \partial \mathcal{M} \simeq \partial \mathcal{X} \neq \emptyset \quad \text{with} \quad \mathcal{M} \subset \mathcal{N}  \quad \quad,
\end{align}
and $\mathcal{X}$ a bounded subspace that encloses all the singularities of the manifold. To be concrete, we can think of this space as $\mathcal{X} \simeq \bigsqcup_i \bar{B}_{r_i} \left( x_i \right) \subset \mathcal{N}$, where  $\bar{B}_{r_i} \left( x_i \right) \neq \emptyset$ are closed balls of radii $r_i$ centered at the point $x_i$. However, $\mathcal{N}$ is only required to be a closed oriented Lorentzian $4$-manifold. Geometrically, this construction allows us to think of the manifold $\mathcal{M}$ as \textit{infinite} with \textit{non-zero (internal) boundary}. It should be noted that this boundary is only important when defining the theory at the level of the action. It ensures 
defining the space-time in which the action and, consequently, the field equations become well-posed. The situation is somewhat illustrated in Figure \ref{symplified}. The physical/observable space-time involved in the action is of course more complex, as discussed below.

\begin{figure*}
\centering
\begin{tikzpicture}[scale=1.5]
\draw [loosely dashdotted, fill=gray!15] plot [smooth cycle] %
    coordinates {(-2.04, 1.54) (-3.52, 2.66) (2.22, 3.22) %
    (4.48, 1.1)(4.44, 0.7) (3.38, 0.98) (0.84, 1.26)};
\draw [dotted, fill=white!15] plot [smooth cycle] %
    coordinates {(3.04, 1.54) (3.52, 1.66) (3.22, 1.22)};
\draw [dotted, fill=white!15] plot [smooth cycle] %
    coordinates {(2, 2.3) (2.5, 2.5) (2.22, 2.1)};
\draw [dotted, fill=white!15] plot [smooth cycle] %
    coordinates {(-1.7, 2.7) (-1.52, 2.4) (-1.22, 2.7)}; 
\draw [dotted, fill=white!15] plot [smooth cycle] %
    coordinates {(-0.04, 2.54) (-0.52, 2.66) (-0.22, 2.22)};
\draw [dotted, fill=white!15] plot [smooth cycle] %
    coordinates {(1, 2.34) (1.6, 2.46) (1.22, 2.02)};    
\draw [dotted, fill=white!15] plot [smooth cycle] %
    coordinates {(-2.04, 2.3) (-2.52, 2.2) (-2.22, 2.5)}; 
\draw [dotted, fill=white!15] plot [smooth cycle] %
    coordinates {(-1.1, 1.6) (-1.3, 2) (-0.8, 1.7)};
\draw[dashed] (2.25,2.3) circle (11pt);
\draw[dashed] (1.27,2.3) circle (12pt);
\draw[dashed] (-0.27,2.5) circle (11pt);
\draw[dashed] (-1.45,2.6) circle (10pt);
\draw[dashed] (-2.3,2.35) circle (10pt);
\draw[dashed] (-1.05,1.85) circle (10pt);
\draw[dashed] (3.25,1.5) circle (11pt);
\draw[dashed] (1.5,1) node {$\mathcal{N}$}; 
\draw[dashed] (0.5,1.5) node {$\mathcal{M}$}; 
\draw[dashed] (0.3,3) node {$ \bar{B}_{r_i} \left( x_i \right)$}; 
\draw (-0.27,2.5) node {$x_i$};
\end{tikzpicture}
\caption{\label{symplified} A two-dimensional representation of the space-time manifold $\mathcal{M}$. The boundary of the manifold is internal and defined by $\partial \mathcal{M} \simeq \partial \mathcal{X}$.}
\end{figure*}

The reason we consider finite volume boundaries around singularities stems from a more realistic examination of the kind of BHs boundaries expected in the Universe. 
Following \cite{LeDelliou:2019mus,Espiro:2019xkd}, those are assumed to be formed, as commented below, from BHs boundaries that can be crudely modeled with Schwarzschild 2-horizons evolving along their semiclassical complete evaporation interval, thus yielding a finite volume excising the problematic regions containing the singularities. In fact, more realistically we consider now $\mathcal{X}=\bigsqcup_{i}\mathcal{B}_{i}\subset\mathcal{N}$, where
$\mathcal{B}_{i}\neq\emptyset$ are the smallest closed set containing
BH $i$'s horizon.

Given the principal bundle structure $P \cong \mathcal{M} \times G$, then $\partial \mathcal{M} \cong \partial P / G$, since the Lie group $G$ is by construction boundaryless. Moreover, if we take $G$ to be the Lie group associated to the isometries of $\mathcal{M}$, it then acts on the fibres via the monodromy action\footnote{In the particular case of Lorentzian manifolds with Levi-Civita connection, this group has been identified with $\text{SO}^0 \left( 1, 3\right)$  \cite[see][and references therein]{10.1007/978-3-642-22842-1_7,Galaev:2015mxf}.
}. As such, the frames are associated to its tangent bundle $T \mathcal{M}$. The latter is somewhat justifiable by the existence of a vierbein. 
In general terms, this gives the picture that the $\partial \mathcal{M}$ surrounds only the \textit{essential singularities} of $P$. We want to associate these singularities to \textit{black-hole}-like elements of the space time $\mathcal{M}$. Topologically speaking, if we account for the proper causal structure of space-time, we can also cover defects like higher-dimensional knots or linked structures by these same Schwarszchild closed balls $\bar{B}_{r_i} \left( x_i \right)$ , in the rationale that they will resist the untangling or contracting operations by continuous deformations of the latter. Causality can then be implemented as proper boundary conditions for the differential forms characterizing the theory. We aim to justify these claims in the following sections.

\subsection{\label{subsec:action} The total action and its field equations.}

The above discussion opens up the possibility of including spinor degrees of freedom in this framework. Locally, we can characterize the latter through a compactly supported Clifford algebra with the following conventions: $\bar{\psi},\psi \in \Omega^0 \left( \mathcal{M} \right) \times \mathbb{G} \simeq S$ with $\mathbb{G}$ being the set of complex Grassmann numbers, $e := \gamma_a \otimes e^a \cong e^a \otimes \gamma_a \cong \gamma_a e^a$ with $\gamma_a$ a representation of the Clifford algebra $\text{Cl}_{1,3} \left( \mathbb{R} \right)$, satisfying $\left\{ \gamma_a , \gamma_b \right\} = 2 \eta_{ab} 1$ (with $\eta_{ab}$ the Minkowskian metric, See Table \ref{tab:VEP-geometrical-elements}) so that $\gamma_b = \eta_{ab} \gamma^b$ along with the existence of $\gamma_* := \frac{i}{4!} \epsilon^{abcd} \gamma_a \gamma_b \gamma_c \gamma_d$, the highest degree \textit{gamma matrix} with $i = \sqrt{-1}$. As it is customary in physics, we take the conjugate transpose of $\psi$ to be $\bar{\psi }:= \psi ^{\dagger} \gamma^{0}$ so the spin norm is given by the expression $\left| \psi \right|^2_s =  \bar{\psi} \psi := \left\langle \psi, \psi \right\rangle_s$ and the expectation value of an operator $\mathcal{O} : S \rightarrow S $ is defined by the expression $\left\langle \mathcal{O} \right\rangle_{\psi} := \left\langle \psi , \mathcal{O} \cdot \psi\right\rangle_s $, where $\cdot$ stands for Clifford multiplication.

With these conventions (See Appendix \ref{app:form} for further differential geometry conventions), the total action we will consider is composed of the following terms:
\begin{align} \label{Bodyaction} 
S \left[ e , \omega , \psi ; \alpha \right] & = S_G \left[ e , \omega ; \alpha \right] + S_F \left[ e , \omega , \psi \right] + S_S \left[ e , \phi \right] + S_I \left[ e , \omega, \phi , j , J , \Upsilon \right] \quad \quad .
\end{align}
The semicolon notation means that $\alpha$ is considered a parameter of the theory and, therefore, will not be part of the variation procedure. In the following, $\kappa$ is the scaled gravitational constant. We have included boundary terms by means of the \textit{inclusion of the boundary operator}  $j^*_{\partial}$ (See App \ref{app:form}) in most of the components to render the variation well-posed in the presence of a non-trivial boundary. The details of each term, as well as their first-order variation, are given below: 
\begin{enumerate}
\item The first term is given by the Holst (gravitation sector) action:
\begin{align}\label{actionG}  
S_G =  - \frac{1}{\kappa} \int_{\mathcal{M}} \omega_{ab} \wedge d_{\omega} \Pi^{ab} + \frac{1}{\kappa} \int_{\partial \mathcal{M}} j^*_{\partial} \left( \omega_{ab} \wedge \Pi^{ab} \right) = \frac{1}{\kappa} \int_{\mathcal{M}} R_{ab} \wedge \Pi^{ab} \quad ,
\end{align} 
where $\Pi^{ab} = \left( \star - \alpha^{-1} \right) \cdot \Sigma^{ab}$ is the Holts $2$-form, with $\Sigma^{ab}$ being the Plebanski $2$-form (See Appendix \ref{app:form}) and $\alpha \in \mathbb{R} \setminus \{ 0 \}$ being the so-called Barbero-Immirzi parameter \cite{Barbero_G__1995,Immirzi_1997,Vyas:2022etz}, which is usually presented in the context of the Holts action \cite{Holst_1996,Rovelli_2011, Bodendorfer:2016uat}. This action is equivalent to the Palatini action in the context of GR, i.e., in the case where $\omega_{ab} \rightarrow \bar{\omega}_{ab}$ (up to a sign) and, therefore, equivalent to the Einstein-Hilbert action \textit{on-shell}. \Jav{For completeness, we will express the latter in tensorial fashion by taking the tetrad components in a coordinate basis $e^a_\mu$ where $g_{\mu\nu} = \eta_{ab}e^a_\mu e^b_\nu$, the spin connection $\omega_\mu^{ab} = -\omega_\mu^{ba}$ and the curvature $R_{\mu\nu}^{ab}(\omega) = \partial_\mu\omega_\nu^{ab} - \partial_\nu\omega_\mu^{ab} + [\omega_\mu^{ab} ,\omega_\nu^{ab} ]$. It follows that we can write the action (\ref{actionG}) in the form:
\[
S_G \sim \int_{\mathcal{M}} d^4x \, e \, e^\mu_a e^\nu_b  \left\{ R_{\mu\nu}^{ab} ( \omega ) - \frac{1}{2\alpha} \epsilon^{ab}_{\ \ cd} R_{\mu\nu}^{cd} ( \omega ) \right\} \quad \text{where} \quad e = \det(e^a_\mu) = \sqrt{-g} \quad .
\]
The first term reproduces the Einstein-Hilbert action ($e^\mu_a e^\nu_b R_{\mu\nu}^{ab} = \mathcal{R}$, the Ricci scalar), while the second is identically null if $\omega$ coincides with the Levi-Civita connection. Continuing with the first order formalism, the variation of (\ref{actionG}) yields:}
\begin{align*}
\kappa \frac{\delta S_G}{\delta e^a} = - \left( \left( * \right) - \frac{1}{\alpha} \right) \cdot R_{ab} \wedge e^b \quad \quad \quad  , \quad \quad \quad \kappa  \frac{\delta S_G}{\delta \omega^{ab}} = - d_{\omega} \Pi_{ab} \quad \quad .
\end{align*}

\item The second term is the massless Dirac action:
\begin{align}\label{actionF}
S_F = \frac{1}{2 \kappa} \left( \int_{\mathcal{M}} \star \left\langle i d^* e \right\rangle_{\psi} + \int_{ \partial \mathcal{M}} j^*_{\partial} \star \left\langle i e \right\rangle_{\psi} \right) = \frac{1}{\kappa} \int_{\mathcal{M}} \left\langle  e \wedge \star i D_{\omega} \right\rangle_{\psi} \quad \quad .
\end{align}
This makes the physical effects over the spinor fields to be of purely geometrical origin. \Jav{As before, we also offer the tensorial version of action (\ref{actionF}) for completeness, given by: 
\[ S_F  \sim \int_{\mathcal{M}} d^{4}x \,  e \, \bar{\psi} i{{\nabla }\!\!\!\!/} \psi \quad \quad  \text{where} \quad  \quad i{{\nabla }\!\!\!\!/} \psi = i \gamma^{\mu} \nabla_{\mu} \psi \quad \quad .
\]
Here, $\nabla_{\mu}$ is the covariant derivative associated to the total spin connection $\omega_{\mu} = \omega^{ab}_{\mu} e_a^\nu e_b^\rho \partial_\nu \partial_\rho = \Gamma^{\nu\rho}_{\;\;\;\mu} \,  \partial_\nu \partial_\rho $. Back to the first order formalism, the} variation of (\ref{actionF}) gives the usual field equations:
\begin{align*} 
\kappa  \frac{\delta S_F}{\delta e^a} = \star \left\langle \gamma_ a i D_{\omega} \right\rangle_{\psi} \quad , \quad  \kappa  \frac{\delta S_F}{\delta \omega^{ab}} = \frac{i}{4} \star \left\langle e \sigma_{ab} \right\rangle_{\psi} \quad , \quad \kappa \frac{\delta S_F}{\delta \bar{\psi}} = e \wedge \star i D_{\omega} \cdot \psi \quad , 
\end{align*}
and their hermitian conjugates. Here, we have written $\sigma^{a b}:= i/2 \left( \gamma^a \gamma^b - \gamma^b \gamma^a \right)$ for short and $i D_{\omega}$ is a Dirac operator associated to the total (Ehresmann) connection $\omega$, satisfying $\overline{D_{\omega} \psi} = D_{\omega} \bar{\psi}$. It should be understood as including a minimal coupling in its definition. \Jav{We include this massless term to foster a chiral symmetry breaking mechanism in the model, which results in interesting physical consequences. Notably, it will be seen that a mass of topological origin associated to spinors is induced in the conditions of this model.}

\item The third term is the Dirichlet energy action for the massless scalar field $\phi$, given by:
\begin{align}\label{actionS}
S_S &:= \frac{1}{2 \kappa} \left( \int_{\mathcal{M}} \star \left\langle \phi , \Delta \phi \right\rangle -  \int_{\partial \mathcal{M}} j^*_{\partial} \left( \phi \star d \phi \right) \right) = - \frac{1}{2 \kappa}  \left\| d \phi \right\|^2_g \quad \quad .
\end{align}
As in the case of the spinor fields, any physical effects will be attributed to geometry. \Jav{This action is 
complemented with interaction terms that will be considered separately below. 
Moreover, its tensorial form is given by the massless Klein-Gordon action:
\[
S_S \sim - \frac{1}{2} \int_{\mathcal{M}} d^{4}x \, e \, g^{\mu \nu} \, \nabla_{\mu} \phi \nabla_{\nu} \phi \quad \quad . 
\]}
The first order variation of action (\ref{actionS}) yields the following expressions:
\begin{align*}
\kappa  \frac{\delta S_S}{\delta e^a} = - \frac{1}{2} \left( \left| d \phi \right|^2_g \star e^{\flat}_a + \tau_{ab} \star e^b \right) \quad \quad  , \quad \quad \kappa  \frac{\delta S_S}{\delta \phi} = \star \Delta \phi \quad \quad ,
\end{align*}
\Jav{where the first term comes from the boundary term in \eqref{actionS} and we have identified the \textit{gravitational stress-energy tensor} $\tau_{ab}$ to be:
\begin{align}\label{stressenergy}
\tau_{ab} = - \star \circ \left(  (e_b^{\flat} \wedge \iota_a) + \eta_{ab} \, \text{id} \right) \circ \left( d \phi \wedge \star d \phi \right) = - 2  \left| d \phi \right|^2_g \eta_{ab} \quad \quad ,
\end{align}}
and where $\Delta := \left( d  + d^* \right)^2$ is the Laplace-Beltrami Laplacian with $d$ the exterior derivative and $d^*$ the exterior co-derivative (further details are presented in appendix \ref{app:form}). \Jav{Note that the \textit{gravitational stress-energy tensor} derives from the action variation with respect to the vierbein $e^a$, differently from the usual stress-energy tensor, which uses the metric $g_{\mu\nu} = \eta_{ab}e^a_\mu e^b_\nu$ as a variation parameter.}

\item The fourth component corresponds to the general interaction term:
\Jav{\begin{align}
\nonumber S_I & := \frac{1}{\kappa} \left[ \left( j , \breve{\omega}  \right)  - \frac{1}{2} \left\| J \right\|^2_g + \left( d^* J ,  \phi \right) + \left( J  , d^* \Upsilon \right) - \int_{\partial \mathcal{M}} j^*_{\partial} \left( \phi \star J  + J \wedge \star \Upsilon \right) \right] \\
\label{actionI} & \quad \quad \quad \quad \quad \quad = \frac{1}{\kappa} \left[ \left( j , \breve{\omega}  \right)  - \frac{1}{2} \left\| J \right\|^2_g - \left( J , d \phi \right) - \left( d J  , \Upsilon \right) \right] \quad \quad \quad ,
\end{align}}
where we have defined $\breve{\omega} := \frac{1}{4} \eta^{ab} \omega_{ab}$ for convenience; $j$, $J$ are $1$-forms that we will identify as conserved currents below, and $\Upsilon$ is an auxiliary \Jav{Lagrange multiplier} $2$-form included to be adjusted later. 
\Jav{The tensorial form of this action can be written as:
\[ 
S_I \sim \int_{\mathcal{M}} d^{4}x \, e \, \left\{ j_{\mu} \breve{\omega}^{\mu} - \frac{1}{2} J_{\mu} J^{\mu} - \nabla_{\mu} \phi \, J^{\mu} - \frac{1}{4} \nabla_{\mu} J_{\nu} \Upsilon_{\alpha \beta} \left( g^{\mu \alpha} g^{\nu \beta} - g^{\mu \beta} g^{\nu \alpha} \right) \right\} \quad .\]
where we are writing the Lagrange multiplier $2$-form as $\Upsilon := \frac{1}{2} \Upsilon_{\alpha \beta} \, dx^{\alpha} \wedge dx^{\beta}$. The last term can appear unfamiliar. It 
represents possible antisymmetric interactions between the two second-rank tensors $\nabla_{\mu} J_{\mu}$ and $\Upsilon_{\mu \nu}$.}
We can understand these terms as generalized Lagrange multipliers carrying holonomic constraints. \Jav{Specifically, we aim to adjust the $2$-form multiplier $\Upsilon$ that makes this system consistent.}
The variation of this action yields:
\begin{align*}
\kappa \frac{\delta S_I}{\delta e^a} = \left[ \iota_{j^{\sharp}} \left( \breve{\omega} \right) - \frac{1}{2} \left| J \right|^2_g - L_{J^{\sharp}} \left( \phi \right) - \left\langle d J , \Upsilon \right\rangle \right] \star e^{\flat}_a \quad \quad , \quad \quad \kappa \frac{\delta S_I}{\delta \omega^{ab}} = \frac{\eta_{ab}}{4} \star j \quad \quad , \quad \\
\kappa \frac{\delta S_I}{\delta j} = \star \breve{\omega} \;\;\; , \;\;\; \kappa \frac{\delta S_I}{\delta \phi} = - \star d^* J \;\;\; , \;\;\; \kappa \frac{\delta S_I}{\delta J} = - \star \left( d \phi + d^* \Upsilon - J \right) \;\;\; , \;\;\; \kappa \frac{\delta S_I}{\delta \Upsilon} = \star d J \;\;\; . 
\end{align*}
\end{enumerate}
By defining the operator $\tilde{\alpha} \; \triangleright	:= \left( \left( * \right) - \frac{1}{\alpha} \right) \cdot$ which acts on two Lie degrees of freedom and collecting all corresponding variations to first order to be identically null, we arrive at the following set of field equations:
\Jav{\begin{align}
\label{fe1} & \quad \quad \quad \quad \left( \tilde{\alpha} \triangleright R_{ab} \right) \wedge e^b - \star \left\langle \gamma_ a i D_{\omega} \right\rangle_{\psi} + \mathcal{V} \left( \phi , J , \Upsilon , j , \breve{\omega}  \right) \star e^{\flat}_a = - \frac{1}{2} \left| d \phi \right|^2_g  \star e^{\flat}_a \quad \quad , \\
\nonumber & \quad \quad \quad \quad \quad \text{with} \quad \quad \mathcal{V} \left( \phi , J , \Upsilon , j , \breve{\omega}  \right) :=  L_{J^{\sharp}} \left( \phi \right) + \frac{1}{2} \left| J \right|^2_g + \left\langle d J , \Upsilon \right\rangle - \iota_{j^{\sharp}} \left( \breve{\omega} \right) \quad \quad ; \\
\label{fe2} & \quad \quad \quad \quad d_{\omega} \Pi_{ab} = \frac{i}{4} \star \left\langle e \sigma_{ab} \right\rangle_{\psi} + \frac{1}{4} \eta_{ab} \star j \quad \quad  ; \quad \quad 0 = i D_{\omega} \psi \quad \quad ; \quad \quad 0 = \breve{\omega} \quad \quad ;  \\
\label{fe4} & \quad \quad \quad \quad \quad \quad 0 = \Delta \phi - d^* J \quad \quad ;  \quad \quad  0 = d \phi + d^* \Upsilon - J \quad \quad ; \quad \quad 0 = dJ \quad \quad ,
\end{align}
where, in \eqref{fe1}, we have used the expression \eqref{stressenergy} to simplify its right hand side. We will build our model, starting from the set of equations \eqref{fe4} which, as we will see in the next section, can be interpreted 
as a topological
, as well as a gauge
, term.}

Before delving into solving this system of field equations, we find it convenient to define the following pairing: Let $X \in T P$ and $\nu$ a $p$-form with values in the Lie algebra. Then, we define the operator $\hat{O}_{\nu} \in \text{End} \left( TP \right)$, induced by the operator $O \in \text{End} \left( \Omega^p \left( T^* P \right) \right)$ and the $p$-form $\nu$, through the pairing:
\begin{align}\label{pairing}
\iota_{X} \left( O \cdot \nu \right) := \iota_{\hat{O}_{\nu} \cdot X} \left( \nu \right) 
\;\;\; \text{such that} \;\;\;
\widehat{\left( O^1 O^2 \right)}_{\nu} = \hat{O}^2_{\nu} \hat{O}^1_{\nu} \;\; / \;\; O^1 , O^2 \in \text{End} \left( \Omega^p \left( T^* P \right) \right) \;\;\; .
\end{align}
The latter is a linear operation.  Moreover, we extend this operation to be linear with respect to the spinor inner product, i.e. $\hat{O}_{\nu} \cdot \left\langle \varpi \right\rangle_{\psi} = \langle \hat{O}_{\nu} \cdot \varpi \rangle_{\psi}$ for $\varpi$ a $1$-form with .

The next sections are discussing each of the equations of the system (\ref{fe1}--\ref{fe4}), following a reversed order: Sec.~\ref{sec:scalarfield} will establish how the current in Eq.~\eqref{fe4} is related to the scalar field. Sec.~\ref{sec:spinconnection} tackles the spin connection equation (Eq.~\ref{fe2}a), then interprets Eq.~(\ref{fe2}b) as a parallel spinor condition. Sec.~\ref{sec:curvature} then gathers the information from all fields to insert them into the Einstein's Eq.~\eqref{fe1}.

\subsection{\label{summary:sec2} Summary of the section}

The model is built on a spacetime manifold $\mathcal{M}$ that is not simply connected. It contains internal boundaries $\partial \mathcal{M}$ representing regions excised around singularities, like the interiors of semi-classical black holes. This gives the spacetime non-trivial topology, characterized by non-contractible $3$-dimensional \textit{holes} for which we have written the action and presented the field equations. \Jav{Even though we will continue in differential forms notation, we 
write the total action in tensor notation for comparison with other theories:
\begin{align*}
    S \sim \int_{\mathcal{M}} d^4x \, \sqrt{-g} \left[  \left( \mathcal{R} - \frac{1}{\alpha} * \mathcal{R} \right) + \bar{\psi} i{{\nabla }\!\!\!\!/} \psi - \frac{1}{2} g^{\mu \nu} \, \nabla_{\mu} \phi \nabla_{\nu} \phi  
    + j_{\mu} \breve{\omega}^{\mu} - \frac{1}{2} J_{\mu} J^{\mu} - \nabla_{\mu} \phi \, J^{\mu}   \quad \right. \\ 
    \left. - \frac{1}{4} \nabla_{\mu} J_{\nu} \Upsilon_{\alpha \beta} \left( g^{\mu \alpha} g^{\nu \beta} - g^{\mu \beta} g^{\nu \alpha} \right) \right] \quad \quad \text{where} \quad \quad * \mathcal{R} := \frac{1}{2} \epsilon^{\mu\nu}_{\ \ \alpha\beta} R^{\alpha\beta}_{\ \ \mu\nu} \quad \quad . \quad \quad 
\end{align*}
Recall that this theory must be complemented with the proper boundary conditions consistent with those discussed in the previous section.}

\section{\label{sec:scalarfield} Scalar field and spacetime topology.}

We establish here how the theory presented encodes topological information. 

Let us focus on the set of equations (\ref{fe4}) since we will interpret these as defining a short Hodge decomposition for the $1$-form $J$. The rationale goes as follows: given that $\mathcal{M}$ inherits the oriented-ness, compact-ness and path-connected-ness of $\mathcal{N}$, the \textit{Hodge orthogonal decomposition} for $k$-forms ($0 \leq k \leq 4$) holds: $ \Omega^k \left( \mathcal{M} \right) \simeq E^{k}_D \left( \mathcal{M} \right) \oplus cE^{k}_N \left( \mathcal{M} \right) \oplus CcC^k \left( \mathcal{M} \right)$, where $D$ and $N$ stand for Dirichlet and Neumann boundary conditions, respectively \cite{Cappell}. Here $E^k_D \left( \mathcal{M} \right)$ is the set of exact forms of order $k$, $cE^k_N \left( \mathcal{M} \right)$ is the set of co-exact forms of order $k$ and $CcC^k \left( \mathcal{M} \right) $ is the set of closed and co-closed forms of order $k$ (simply called \textit{harmonic forms} if the boundary is trivial)\footnote{See appendix \ref{app:form} for more information}. This implies that we can always write $J$ as:
\begin{align}\label{hodgedecomp}
J = d \phi + d^* \Phi + \theta \quad \text{satisfying} \quad d J = d d^* \Phi  \quad ; \quad d^* J  = \Delta \phi \quad ; \quad  0 = d \theta = d^* \theta \quad \quad ,
\end{align}
where $d \phi \in E^1_D \left( \mathcal{M} \right)$, $d^* \Phi \in cE^1_N \left( \mathcal{M} \right)$ and $\theta \in CcC^1 \left( \mathcal{M} \right)$, orthogonal to each other. \Jav{Eqs.~\eqref{hodgedecomp} can be understood analogously as the electric field potential $1$-form. In electromagnetism, the first term $d\phi$ would correspond to the electric potential, the second $d^* \Phi$ to the magnetic potential, while the third term $\theta$ is null in that case because of topological reasons: the \textit{space of loops} is contractible in three dimensions when there are no boundaries. Ours is not the case.} Notice that the third equation in (\ref{hodgedecomp}) is the first equation of the set (\ref{fe4}). Taking identically $d d^* \Upsilon = 0$ satisfies the third equation of the set. This can be easily achieved by appropriate Dirichlet boundary conditions on the $2$-form $\Upsilon$. Instead of taking this route, we will consider $d^* \Upsilon \cong \theta$ and present this in the equivalent formulation given by the short Hodge decomposition:
\begin{align}\label{hodgedecompshort}
J = \theta + d \phi \quad \text{where} \quad \quad \theta \cong d^* \Upsilon \quad \quad \text{(defined up to homotopy)} \quad \quad .    
\end{align}
\Jav{Theoretically, we have two 
ways to understand this construction. }
\begin{enumerate}
    \item \Jav{On one hand,} we would like to interpret $J$ as a physical current, which makes the field equation $d^*J = \Delta \phi$ to be interpreted as some sort of anomaly of this theory. We will briefly discuss this in the closing section of the paper. We will dedicate the rest of the section to elaborating on the cohomological implications throughout this section, along with a discussion on boundary values for $\theta$.
    \item \Jav{On the other hand, the decomposition (\ref{hodgedecompshort}) can also be interpreted as gauge transformation of $\theta$ under some abelian Lie group represented by $\phi$ ($J_{\mu} =  \theta_{\mu} + \partial_{\mu} \phi$ in more familiar notation). This makes $\theta$ a gauge field that should have an associated well defined covariant derivative (of the sort $D_\mu \psi = (\partial_\mu - i J_\mu) \psi$). Notice that the massless Dirac Action \eqref{actionF} is explicitly gauge-invariant in this sense but there is no gauge field kinetic term $\sim -\frac{1}{4} F_{\mu\nu} F^{\mu\nu}$ in the total action. The explicit gauge symmetry breaking of the interaction term $\sim - \frac{1}{2} J_{\mu} J^{\mu}$ is then expected to be compensated by the $2$-form multiplier $\Upsilon$. We will show in the following sections that this interpretation indeed can be properly formalized (See \ref{subsec:contandtor} for further elaboration).}
\end{enumerate}

For the time being, \Jav{we can make use of the orthogonality of the short Hodge decomposition (\ref{hodgedecompshort}) to write $L_{J^{\sharp}} \left( \phi \right) = \left\langle J , d \phi \right\rangle = \left| d \phi \right|^2_g $ along with} the Lagrange multiplier's equations into the set (\ref{fe1} -\ref{fe4}), to get the simplified system:
\Jav{\begin{align}
\label{fe11} \left( \tilde{\alpha} \, \triangleright R_{ab} \right) \wedge e^b + \frac{1}{2} \left| J \right|^2_g \star e^{\flat}_a = - \frac{3}{2} \left| d \phi \right|^2_g  \star e^{\flat}_a \quad \quad ; \quad \quad  \\
\label{fe21} d_{\omega} \Pi_{ab} = \frac{i}{4} \star \left\langle e \sigma_{ab} \right\rangle_{\psi} - \frac{1}{4} \eta_{ab} \star j \quad \quad ; \quad \quad 0 = i D_{\omega} \psi \quad \quad ,  
\end{align}
where $J$ has been defined in (\ref{hodgedecompshort}). The tensorial versions of these equations are not particularly illustrative, so we prefer to omit them at this point. Here, the} simplified form of equations \eqref{fe11} and \eqref{fe21} summarize the spirit of the model and the dynamics of its fundamental generalized fields $\left\{ e^a , \omega^{ab} , \phi , J , \psi \right\}$. Equation (\ref{fe11}) is a generalized Einstein equation in the language of differential forms. The leftmost equation in the set (\ref{fe21}) is a generalized spin connection equation, while the rightmost equation is a massless Dirac equation in curved space. Before continuing, some technical remarks about the $\theta$ $1$-form are in order.

\subsection{\label{subsec:theta} Co-homology and boundary value problems.}

Since the manifold $\mathcal{M}$ has a non-trivial boundary $\partial \mathcal{M}$ , we have a further decomposition  of the closed-co-closed sector in (\ref{hodgedecomp}) as follows:
\begin{align}\label{CcCdecomp}
CcC^k \simeq CcC^k_N \oplus EcC^k \simeq CcE^k \oplus CcC^k_D \quad \text{so that} \quad 
\begin{cases}
CcC^k_N \simeq H^k \left( \mathcal{M} \right) \quad \;\; \text{and} \\
CcC^k_D \simeq H^k \left( \mathcal{M} , \partial \mathcal{M} \right)
\end{cases} , 
\end{align}
where $H^k \left( \mathcal{M} , \partial \mathcal{M} \right)$  is the cohomology relative to the boundary $\partial \mathcal{M}$ . Thus, $\theta$ becomes a representative of the equivalence class $\left[ \theta \right]$ (under homotopy) associated to an element of the co-homology group $H^1 \left( \mathcal{M} \right)$ if strict Neumann boundary conditions are considered. However, we must recall that by the Hodge–Weyl theorem, every De Rham cohomology class has a \textit{unique harmonic representative}. Summing up, $\theta$ is unique up to homotopy such that \Jav{$j^*_{\partial} \left( \star \theta \right) \ne 0$}. In other words, the relevant topological information contained in the chain of isomorphism of equation (\ref{homocohomo}) is essentially encoded in the $1$-form $\theta$ if Neumann boundary conditions are assumed, which we will do from now on. Notice that this makes $J$ also an element of the homotopy class $[\theta]$ (this is usually written $J \sim \theta$).

The above picture is completely consistent with the remark on section \ref{sec:structure} that the non-contractible $ 3$-volumes can naturally measure the essential topological features of this construction. Technically, we are interested in the $\pi_3 \left( \mathcal{M} \right)$ homotopy group. Being so, the following chain of isomorphisms is then justified: 
\begin{align}\label{homocohomo}
\pi_3 \left( \mathcal{M} \right) 
 \simeq H_3 \left( \mathcal{M} \right) \simeq H_3 \left( \mathcal{N} , \mathcal{X} \right) \simeq H^1 \left( \mathcal{N} \setminus \mathcal{X} \right) \simeq H^1 \left( \mathcal{M} \right) \quad \quad ,
\end{align}
where we have used the \textit{Hurewicz theorem}, the \textit{excision} theorem and the \textit{Poincar\'{e}-Lefschetz duality} \cite{hatcher}. As such, the rank of $H^1 \left( \mathcal{M} \right)$ will be equal to that of $H_3 \left( \mathcal{M} \right)$ and can be directly associated with the problem of calculating the Betti number $b_3$ of $\mathcal{M}$. That is, we are interested in the $3$-holes of the manifold $\mathcal{M}$. Notice that this reasoning is valid for the groups $H^1 \left( \mathcal{M} \right)$ and $H_3 \left( \mathcal{M} \right)$ which are not depending on any of the other sets defined in (\ref{regiondef}). Although we have defined these in the base space $\mathcal{M}$, the results are consistent with the principal bundle $P$ when canonically identifying $H^i \left( P \right) \cong H^i \left( \mathcal{M} \right)$ via the isomorphism induced by the projection map $\pi: P \rightarrow \mathcal{M}$.

\subsection{\label{subsec:BVproblem} The boundary Value Problem and " theoretical topological probes".}

Following \cite{BELISHEV2008128}, by the Friedrichs decomposition, we can pose the problem of determining the harmonic $1$-form $\theta$ as a Boundary Value (hereafter B.V.) problem for a scalar field. Note that the harmonic $1$-form can be interpreted as the representative of all possible non-contractible $3$-volumes from Eq.~\eqref{homocohomo}. Let $\varphi \in \Omega^0 \left( \partial \mathcal{M} \right)$ such that $\theta = d \varphi$, be the solution of the following B.V. problem: 
\begin{align}\label{BVproblem}
\left\{ 
\begin{array}{cc}
\Delta \varphi = 0 ,& d^* \varphi = 0 \\
j^*_{\partial} \left( \varphi \right) =  \xi &
\end{array}
\right. \quad \quad \text{for some} \quad \quad \xi \in \Omega^0 \left( \partial \mathcal{M} \right) \quad \quad ,
\end{align}
where the solution $\varphi$ is unique up to an arbitrary Dirichlet harmonic field. More importantly, the \textit{Dirichlet to Neumann operator} (DN) $\Lambda: \Omega^0 \left( \partial \mathcal{M} \right) \rightarrow \Omega^3 \left( \partial \mathcal{M} \right)$  defined by the expression $\Lambda \xi = j^*_{\partial} \left( \star d \varphi \right) = j^*_{\partial} \left( \star \theta \right)$ is a well-defined operator since it turns out to be independent of the choice of the solution $\varphi$.\Jav{ Under these conditions, the following equation $ d \left( \varphi \star \theta \right) = \star \left| \theta \right|^2_g $ is satisfied locally, since $\Delta  \varphi = 0$, and we can write:
\Jav{\begin{align*}
\int_{\partial \mathcal{M}} j^*_{\partial} \left( \varphi \star \theta \right)  = \int_{\partial \mathcal{M}} \xi \Lambda \xi 
= \int_{\mathcal{M}} \star \left| \theta \right|^2_g 
\quad \;\; \Rightarrow  \quad \;\; d \left( \xi \Lambda \xi \right)  - \star \left| \theta \right|^2_g = 0 \quad \;\; \text{locally} \quad \;\; .
\end{align*}}
The latter is equivalent to
\begin{align}\label{DNexpression}
\left| \theta \right|^2_g 
= d^* \Lambda_{\xi} \quad \quad \text{where we have defined} \quad \quad \Lambda_{\xi} := \star \left( \xi \Lambda \xi \right) \quad \quad ,
\end{align}
which we will call the \textit{cosmological $1$-form}, that only depends on boundary conditions.} Furthermore, from a strict B.V. problem context we can calculate the third Betti number using the expression $b_3 = \text{dim} \, \text{Ran} \left[ \Lambda + d \Lambda^{-1} d  \right]$ \cite[see][and references therein for further details]{BELISHEV2008128}. In other words, we can think of $\varphi$ as being a \textit{theoretical topological probe}, since by studying the B.V. problem (\ref{BVproblem}) for the scalar field $\varphi$ we will find the relevant topological information of these models. This problem can be studied under cosmological settings if the harmonic $1$-form potential's boundary value $\xi$, analogous to BHs quasinormal modes, and $\partial \mathcal{M}$ are akin to observations, a problem we would like to elucidate in ongoing investigations, but that remains out of the scope of this present paper. 
We shall instead focus on explicitly sketching a possible mechanism for how this topological information can enter the dynamics of a gravitational theory. This is introduced in the following section.

\subsection{\label{summary:sec3} Summary of the section}

In this section, the generalised Einstein Field Equations coupled to its Dirac and Connection equations allow to encode the key topological information of the model in a harmonic 1-form $\theta$ (closed and co-closed). This form is a unique representative of a cohomology class $H^1 \left( \mathcal{M} \right)$, which can be associated to the $3$-holes measured by the homotopy group $\pi_3 \left( \mathcal{M} \right)$. This way, the topology dynamics are controlled by the Black Holes (3-holes) within the spacetime.

\Jav{The simplified system of field Eqs.~(\ref{fe11},\ref{fe21}) used by this section  in tensorial form are cumbersome and not particularly enlightening so we will omit them.}

\section{\label{sec:spinconnection} Spin connection and Contortion.}

In this section we will build the curvature and torsion generating connection that reflects the topological information clarified above. 

Let us examine the simplified set of field equations (\ref{fe21}). It is important to note that, if $\theta$ is identified as representative of a \textit{harmonic $1$-form}, its presence in the field equations must be properly justified in terms of the $G$-bundle structure. Initially, as a solving strategy, we will consider the usual decomposition of the total connection $\omega = \bar{\omega} + K$, where $\bar{\omega}$ is the (torsion free) Levi-Civita connection, while $K$ is the contortion $1$-form, such that:
\begin{align} \label{connection} 
d_{\omega} e^a := T^a = K^{a}_{\;\;b} \wedge e^b \; \in \; \Omega^2 \left( \mathcal{M} \right) \quad \quad  \text{where $T^a$ is the torsion $2$-form.} \quad \quad . 
\end{align}
Furthermore, Eq. (\ref{connection}) can be inverted by means of the interior product\footnote{By abuse of notation, sometimes we will write the analogous operation induced by the metric as $\iota^a := \eta^{-1}_{ab} \iota_b := \eta^{ab} \iota_b $.} to yield:
\begin{equation}\label{contortion1}
{K_{ab}} =  - \frac{1}{2}\left\{ {{\iota_a}\left( {{T_b}} \right) - {\iota_b}\left( {{T_a}} \right) - {\iota_{a \wedge b}}\left( {{T_c}} \right) \wedge {e^c}} \right\} \quad \quad,
\end{equation}
so the relation is one-to-one. Given the underlying principal $G$-bundle structure we are working with, the contortion can be taken as a $G$-connection and consequently interpreted as an \textit{Ehresmann connection} or, equivalently, it can be taken to act solely on the horizontal bundle (note that the horizontal part of the bundle is the kernel of the connection $1$-form). All these will become consistent in the following subsection.

The curvature $2$-form associated to the total connection $\omega$ is given by the expression $R^a_{\;\;b} = d \omega^a_{\;\;b} + \omega^a_{\;\; c} \wedge \omega^c_{\;\;b} \; \in \, \Omega^2\left(\mathcal{M} \right) $ , so that the decomposition above Eq.~\eqref{connection} implies the associated curvature decomposition:
\begin{align}\label{curvaturedecomp}
R^a_{\;\;b}  
= \left( d \bar{\omega}^a_{\;\;b} + \bar{\omega}^a_{\;\; c} \wedge \bar{\omega}^c_{\;\;b}\right) +\left( d_{\bar{\omega}} K^a_{\;\;b} + K^a_{\;\; c} \wedge K^c_{\;\;b}\right)
 := \bar{R}^a_{\;\;b} + F^a_{\;\;b} \quad \quad ,
\end{align}
where $\bar{R}^a_{\;\;b}$ is recognized as the torsion-less (Levi-Civita) curvature $2$-form, formally equivalent to that of a classical GR curvature $2$-form, while the remaining term $F^a_{\;\;b}$, referred subsequently as the \textit{field strength $2$-form}, concentrates all the contributions from the contortion \cite{Fischer:2020lri}. To complete the picture, given that \textit{the second Bianchi identity} (See \ref{Bianchi}) is always satisfied for any connection, it is sufficient to focus on the \textit{first Bianchi identity} (Also in \ref{Bianchi}). In this context, a simple calculation shows that:
\begin{align}\label{Bianchi2}
d_{\omega} T^a = d_{\omega} K^a_{\;\;b} \wedge e^b - K^a_{\;\; b} \wedge T^b 
= F^a_{\;\; b} \wedge e^b \quad \quad \text{this is} \quad \quad \bar{R}_{ab} \wedge e^b = 0  \quad \quad ,
\end{align}
as expected.

We can argue that since the model presented in subsection \ref{subsec:action} restricts the bilinear $\left\langle i e \right\rangle_{\psi}$ to be \textit{bosonized} by being identified with the $1$-form $J$, we are not expecting terms that would couple to the torsion sector as an additive spinor contribution to the total connection. In other words, only contributions from elements of the homotopy class $[\theta]$ are expected. For definiteness, take some general $1$-form in the class $[\theta]$, so that the contortion $K_{ab}$ is dependent on it. Inserting this back via the connection decomposition\footnote{Risking a bit of sloppiness, we use a similar notation for the $1$-form $\omega$ and its representation over the spinor bundle, which should be clear in context.} $\omega = \bar{\omega} + K$ and equation (\ref{connection}) into the first equation of set \eqref{fe21} we get:
\begin{align} 
\label{fe221} \frac{1}{2} \tilde{\alpha} \, \triangleright \left( T^a \wedge e^b - e^a \wedge T^b \right) = \star \left( \frac{i}{4} \left\langle e \sigma^{ab} \right\rangle_{\psi} + \frac{1}{4} \eta_{ab} J \right) \quad \quad ,
\end{align}
In this section, we will focus on finding a suitable contortion $K_{ab}$ that is consistent with the mathematical structure presented and will help to solve the field equations.

\subsection{\label{subsec:contandtor} General form of the Contortion and Torsion.}

To obtain the general form of the contortion and torsion, we will first obtain the Clifford basis representation of $K_{ab}$ which depends on a representative of the class $[\theta]$ in the spinor bundle, say $J$ for definiteness. Let the Lie algebra of $\text{Cl}_{1,3} \left( \mathbb{R} \right)$ be denoted by $\mathfrak{cl}_{1,3} \left( \mathbb{R} \right)$, so that we can construct the group $G_{[\theta]}$  \cite[further technical details can be seen in][]{agricola2003holonomy} with its corresponding Lie algebra $\mathfrak{g}_{[\theta]}$ associated to the locally 
Lorentz transformations of $\mathbb{R}^{1,3}$ preserving the closed and co-closed class $[\theta]$. If we denote by $\mathfrak{g}^*_{[\theta]}$ the subalgebra of $\mathfrak{cl}_{1,3} \left( \mathbb{R} \right)$ generated by elements of the form $i \iota_{e^{\sharp}} \left( J \right)$, where 
$\iota : \Omega^k \left( \mathcal{M} \right) \rightarrow \Omega^{k-1} \left( \mathcal{M} \right)$ is the interior product and $e^{\sharp} = \gamma^a e_a$, where $\left\{ e_a \right\}$ is the frame basis, then $\mathfrak{g}^*_{[\theta]}$ is invariant under the isotropy group $G_{[\theta]}$. Moreover, since $J$ is a $1$-form, the \textit{infinitesimal holonomy algebra of $[\theta]$}, denoted $\mathfrak{h}^*_{[\theta]} = \left[ \mathfrak{g}^*_{[\theta]} , \mathfrak{g}^*_{[\theta]}\right] \subset \mathfrak{cl}_{1,3} \left( \mathbb{R} \right)$, is known to be a \textit{compact Lie group}. Additionally, since $\mathfrak{h}^*_{[\theta]}$ is a Lie subalgebra of $\mathfrak{cl}_{1,3} \left( \mathbb{R} \right)$, there is a unique connected immersed Lie subgroup $H^*_{[\theta]} \subseteq \text{Cl}_{1,3} \left( \mathbb{R} \right)$ whose Lie algebra corresponds to $\mathfrak{h}^*_{[\theta]}$. Finally, since the representative of the class $[\theta]$ is unique by the Hodge-Weyl theorem, this Lie algebra is one-dimensional and consequently Abelian. In other words, $H^*_{[\theta]}$ performs abelian gauge transformations.

Now, to obtain the Clifford representation of the contortion, we can start by the adjoint representation of the Lie algebra $\mathfrak{h}^*_{[\theta]}$ acting on a general Clifford vector $e^{\sharp} = \gamma^a e_a$, this yields:
\begin{align*}
\text{Ad}_{J} \left( e^{\sharp} \right) = \left[ J , e^{\sharp} \right] = k_{ab} \left[ \gamma^a , \gamma^b \right] \quad \quad  \text{where} \quad \quad k_{ab} := \frac{1}{2} \left( \iota_a \left( J \right) e_b - \iota_b \left( J \right) e_a  \right) \quad \quad .
\end{align*}
Notice that $k_{ab}$ defines a $1$-form through the application of the $\flat$ musical isomorphism:
\begin{align}\label{defk}
k_{ab}^{\flat} \left( J \right) = \frac{1}{2}\left( \iota_a \left( J \right) e^{\flat}_b - \iota_b \left( J \right) e^{\flat}_a \right) = \iota_{J^{\sharp}} \left( \Sigma^{\flat}_{ab} \right) \quad \quad \text{(co-adjoint representation)} \quad \quad , 
\end{align}
which can subsequently be interpreted as an affine connection. However, since $J$ is a $1$-form, all endomorphisms $\iota_{e^{\sharp}} \left( J \right)$ acting on spinor elements are skew-symmetric. Hence, a general connection associated to the $1$-form $J$ should be given by the expression\footnote{We are noting by $\left( * \right) A_{ab} = A^{\left( * \right)}_{ab} : = \frac{1}{2} \epsilon_{abcd} A^{cd}$ the Lie dual acting over any $A^{cd} \in \Omega \left( \mathcal{M} \right)$ with two spin indices $c,d$ and by $\star A_{ab}$ the hodge dual of the form $A_{ab}$. The latter is defined over the vierbein basis as $\star \left( e^{a_1} \wedge \cdots \wedge e^{a_p} \right) := \frac{1}{\left( n - p\right)!} \epsilon^{a_1 \cdots a_p}_{\;\;\;\;\;\; \;\;\;\;\;a_{n+1} \cdots a_n} e^{a_{p+1}} \wedge \cdots e^m$,  $( p < n)$ and it extends to $\Omega \left( \mathcal{M} \right)$ by linearity.}:
\begin{align}\label{quasicontorion}
K_{ab} \sim  i B^{-1} \frac{\left( 1 + A \left( * \right) \right)}{2} \cdot k_{ab}^{\flat} = i \iota_{\bar{J}^{\sharp}} \left( \frac{1}{2} \left( 1 + A \, \star \right) \cdot \Sigma^{\flat}_{ab} \right) \;\;\; \text{for some $A,B \neq 0$} \;\;\; ,
\end{align}
where we have inserted back the imaginary $i$ coming from the generators of the Lie algebra $\mathfrak{h}^*_{[\theta]}$. Notice that, in this case, the constant $B$ is associated to the scale invariance,  since $B^{-1} \iota_{J^{\sharp}} \left( \cdot \right) = \iota_{ B^{-1} J^{\sharp}} \left( \cdot \right)$, which has been obtained solely based on linear independence. For the moment, since $J$ is defined up to homotopy invariance, we can redefine $J \sim \bar{J} := B^{-1} J$. Nevertheless, $B$ is constrained by dimensional consistency, while the constant $A$ must be dimensionless and is responsible for the Lie symmetry of the term $K_{ab}$. We postpone a more physical discussion to subsequent sections. Summing up, we have:
\begin{align}\label{contandtor} 
K_{ab} = i \, \iota_{\frac{1}{2} \widehat{\left( 1 + A \, \star \right)}_{\Sigma} \cdot \bar{J}^{\sharp}} \left( \Sigma^{\flat}_{ab} \right) \quad \;\; \text{and} \;\; \quad T_a = K_{ab} \wedge e^b = \frac{i}{2} \left( \star - \frac{A}{2} \right) \cdot \left( e^a \wedge \bar{J} \right) \quad \quad , 
\end{align}
as local forms for the contortion (\ref{quasicontorion}) and its associated torsion, respectively.
For completeness, we also give the general form of the field strength $2$-form :
\begin{align}\label{preupsilon}
F_{ab} = \frac{ i\left( \left( * \right) + A \right)}{4} \cdot \left[ L_a \left( \bar{J} \right) \wedge e^{\flat}_b - L_b \left( \bar{J} \right) \wedge e^{\flat}_a \right] + \star \left[ \left( \frac{1 - A^2}{2} \left( * \right) + A \right) \cdot \frac{i \bar{J}}{2} \wedge k_{ab}^{\flat} \left( \bar{J} \right) \right] \;\;\; , 
\end{align}
where $L_a \left( \, \cdot \, \right)$ is the Lie derivative (See Appendix \ref{app:form}). Here it is explicit that the remaining parameter $A$ is related to the Lie symmetry. Furthermore, we can symmetrize the latter by imposing $\left. K_{ab} \right|_{A = \pm i}$ , making the contortion anti-self-dual/self-dual, respectively. We will make these preferences for reasons that will become apparent in the following paragraphs.

\subsection{\label{subsec:spin} The spin field equation and explicit Lie symmetry}

Let us return to the set of equations (\ref{fe221}) and use equation (\ref{contandtor}) again on the first equation. This will allow us to find additional restrictions for the parameter $A$. After some exterior algebra, we obtain:
\begin{align}\label{fe222}
\frac{\alpha^{-1} i}{4} \eta^{ab} \star \bar{J} + \frac{A i}{2} \, \bar{J} \wedge \Pi^{ab} = \frac{\eta^{ab}}{4} \star j  + \frac{i}{4} \star \left\langle e \sigma^{ab} \right\rangle_{\psi} 
\quad \quad .
\end{align}
Hence:
\begin{enumerate}
\item From the symmetrical part of equation (\ref{fe222}) we get $\alpha^{-1} \bar{J} = - i \, j$ and they immediately satisfy the conserved current condition $dj =d \bar{J} =0$. If we write $\bar{J} := \bar{J}_a e^a$, by using equation (\ref{contandtor}), we can calculate:
\begin{align*}
0 = d \bar{J} = i D_{\omega} \bar{J}_a \wedge e^a + \bar{J}_a T^a = i D_{\omega} \bar{J}_a \wedge e^a + i \left( \star - \frac{i}{2} \right) \cdot \left( \bar{J} \wedge \bar{J} \right) = i D_{\omega} \bar{J}_a \wedge e^a \quad \quad , 
\end{align*}
which forces $i D_{\omega} \bar{J}_a = 0$, i.e. the spinor current $\bar{J}_a$ must be Ehresmann flat. 

\item From the skew symmetrical part of equation (\ref{fe222}), by using equation (\ref{defk}), after \textit{hodge dualizing}, the left side of the skew symmetric part in equation (\ref{fe221}) can be written as:
\begin{align*}
\frac{A i}{2} \star \left( \bar{J} \wedge \Pi^{ab} \right) = - \frac{A i}{2}  \iota_{\bar{J}^{\sharp}} \left( \star \Pi^{ab} \right) = i \iota_{\frac{A}{2} \widehat{\left( 1 + \star \alpha^{-1} \right)}_{\Sigma} \cdot \bar{J}^{\sharp}} \left( \Sigma^{ab} \right) \quad \quad ,
\end{align*}
where we have used the pairing dualization of equation (\ref{pairing}). The right side of the skew-symmetric part in equation (\ref{fe222}), after performing some Clifford and exterior algebra manipulations, gives:
\begin{align*}
\frac{i}{4} \left\langle e \sigma^{ab} \right\rangle_{\psi} 
= \iota_{\langle \hat{O}_{\Sigma} \cdot e^{\sharp} \rangle_{\psi}} \left( \Sigma^{ab} \right) \quad \quad \text{where} \quad \quad O := \mathcal{P}_+ \otimes P_+ + \mathcal{P}_- \otimes P_- \quad \quad,
\end{align*}
and where we have identified the operators\footnote{Also, these projectors induce a decomposition $\Omega^2 \left( \mathcal{M} \right) \cong \mathcal{P}_+ \Omega^2 \left( \mathcal{M} \right) \oplus \mathcal{P}_- \Omega^2 \left( \mathcal{M} \right)$ of self-dual and antiself-dual differential forms and a chiral decomposition of the spinor set $S \cong P_+ S \oplus P_- S$, respectively.}:
\begin{align}\label{chiralproj}
\mathcal{P_{\pm}} := \frac{1}{2} \left( 1 \pm i \star \right) \; \text{($2$-form projector)} \;\; \text{and} \;\; P_{\pm} := \frac{1}{2} \left( 1 \pm \gamma_*\right) \; \text{(chiral projector)} \;\; .
\end{align}
Notice that, thus defined, $O$ also defines a projector operator in $\text{End}\left( \Omega^2 \left( \mathcal{M} \right) \otimes S \right)$. When combining the left and right side of the skew symmetric part of equation (\ref{fe222}) , we arrive at the operational equation:
\begin{align}
\label{skewt1}  & 
\iota_{\bar{J}^{\sharp}_{\alpha, A}} \left( \Sigma^{ab} \right) = \frac{2 \alpha i}{A} \iota_{\langle \left( \frac{P_+}{\alpha - i} + \frac{P_-}{\alpha + i} \right) \cdot e^{\sharp} \rangle_{\psi}} \left( \mathcal{P}_+ \cdot \Sigma^{ab} \right) \quad \quad \text{with} \quad \quad \alpha \neq \pm i \quad \quad ,
\end{align}
where we have used $\left( 1 + \star \alpha^{-1} \right)^{-1} = \frac{\alpha}{1 + \alpha^2} \left( \alpha  - \star \right)$ and the fact that the $2$-form projectors satisfy the expression $\mathcal{P}_{\pm} \cdot \left( \alpha - \star \right) = \left( \alpha \pm i \right) \mathcal{P}_+$. 
\end{enumerate}

Formally speaking, from this last result and using equation (\ref{contandtor}), we obtain the preliminary operational equation for the contortion $1$-form in terms of the parameters $\alpha$ and $A$ as:
\begin{align}\label{skewt2} & K_{ab} 
= \frac{i}{2} \left( 1 - iA \right) \iota_{\bar{J}^{\sharp}_{\alpha, A}} \left( \Sigma^{ab} \right) \quad \quad  .
\end{align}
Notice the presence of the operator $\mathcal{P}_+$ in equation  \eqref{skewt1} that explicitly breaks the symmetry of the contortion $K_{ab}$ in equation (\ref{contandtor}). Relating this last expression with equation (\ref{preupsilon}), we can see that $A$ can be chosen independently of the Barbero-Immirzi parameter $\alpha$. However, the choice $A = -i$ nullifies the contortion and, consequently, the torsion degrees of freedom of the theory. Hence, consistency requires $A := i$, turning $K_{ab}$ into an anti-self-dual connection. 
With this choice, we have:
\begin{align}\label{Jcurrent}
\bar{J} \cong \left. \bar{J}_{\alpha,A} \right|_{A=i} = \alpha \langle \left( \frac{2}{i \alpha + 1} P_+ + \frac{2}{i\alpha - 1} P_- \right) \cdot e \rangle_{\psi} = \alpha j \quad \quad .
\end{align}
This is an effective chiral symmetry breaking meadited by the BI parameter $\alpha$. It can be seen that the choices $\alpha = \pm i$ are somewhat forbidden in this theory, unless they are compensated by allowing the spinor to be specifically chiral, according to the above equation. We will consider this problem in the next section. For the moment, we update the contortion and torsion depending on the current $1$-form $\bar{J}$ to be:
\begin{align}\label{torsion2}
K_{ab} \left( \bar{J} \right) = i \, \iota_{\bar{J}^{\sharp}} \left( \mathcal{P}_+ \cdot \Sigma_{ab}^{\flat} \right) \quad \quad  \text{and} \quad \quad  T^b = i \left( \star - \frac{i}{2} \right) \cdot \left( e^b \wedge \bar{J} \right) \quad \quad ,
\end{align}
where the parameter $\alpha$ is yet free and will be adjusted in the next section. The latter expression also simplifies equation (\ref{preupsilon}) to:
\begin{align}\label{upsilon}
F_{ab} \left( \bar{J} \right) = 
\left[ \mathcal{P}_+ \cdot \left( L_a \left( \bar{J} \right) \wedge e^{\flat}_b - L_b \left( \bar{J} \right) \wedge e^{\flat}_a \right) \right] + \star \left( \bar{J} \wedge K_{ab} \left( \bar{J} \right) \right) 
\quad \quad .
\end{align}


\subsection{\label{summary:sec4} Summary of the section}

We have constructed the model's spacetime bundle connection and decomposed it into the standard Levi-Civita connection (torsion-free) and a contortion $1$-form (responsible for torsion). We have shown that the latter is naturally induced by the existence of the harmonic form $\theta$. Henceforth, we can talk of a topologically induced torsion.

\Jav{The tensorial form of the field strength (curvature) associated with the contortion part of the connection reads as follows. }\Jav{Recall that $F_{ab} = \frac{1}{2}  e^{\mu}_a e^{\nu}_b F_{\mu \nu \alpha \beta} \,  dx^{\alpha} \wedge dx^{\beta}$ where the tensor coefficients are now given by:
\begin{align*}
F_{\mu \nu \alpha \beta} & = \left( g_{\beta \mu} \nabla_{\alpha} \bar{J}_{\nu} - g_{\beta \nu} \nabla_{\alpha} \bar{J}_{\mu} \right) + \frac{i}{2} \epsilon^{\delta \gamma}_{\;\;\; \alpha \beta}  \left( g_{\gamma \mu} \nabla_{\delta} \bar{J}_{\nu} - \frac{1}{2} g_{\gamma \nu} \nabla_{\delta} \bar{J}_{\mu} \right) + \\
& - \frac{i}{4} g_{\zeta \mu} g_{\eta \nu} \bar{J}^{\delta} \left( \bar{J}^{\eta}  \epsilon^{\zeta}_{\;\; \delta \alpha \beta} - \bar{J}^{\zeta} \epsilon^{\eta}_{\;\; \delta \alpha \beta} \right) + \frac{1}{4} \epsilon^{\delta \gamma}_{\;\;\; \zeta \iota} g_{\gamma \mu} g_{\delta \nu} \bar{J}^{\eta} \left( \bar{J}^{\zeta} \epsilon^{\iota}_{\;\; \eta \alpha \beta} - \bar{J}^{\iota} \epsilon^{\zeta}_{\;\; \eta \alpha \beta} \right) \quad \quad .
\end{align*}}

\section{\label{sec:parallelspinors} Ehresmann parallel spinor hypothesis}

In this section, we connect the torsion with the topology of the holes with a spinor field described with a gauge theory-like geometry. 

\begin{figure}[h!]
\centering
\includegraphics[width=0.55\textwidth]{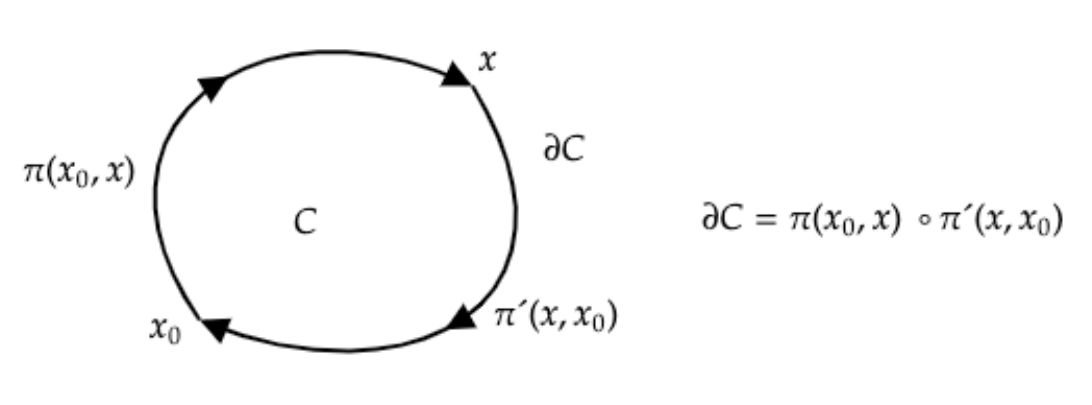}
\caption{\label{holonomyfig} A typical closed parallel transport $\partial C$ measuring the holonomy with base point $x_0$ of the total connection $\text{Hol}_{x_0} \left( \bar{\omega} \right)$.}
\end{figure}

The massless Dirac field equation (\ref{fe2}) (or equivalently, (\ref{fe21})) plus consistency with the symmetrical part of equation (\ref{fe222}), pushes for a purely geometrical \textit{parallel spinor} interpretation with respect to the (total) Ehresmann connection $\omega$. We call this condition an \textit{hypothesis}, however, we will assume it to be true from now on\footnote{In the context of Lorentzian manifolds with Levi-Civita connections (no torsion), parallel spinors have been studied in \cite{Moroianu_2004,alexandrov2005eigenvalue} and the references therein. In contrast, many of the idiosyncrasies of the torsion version, which is the one we are interested in, have been studied in \cite{agricola2013twistorialeigenvalueestimatesgeneralized,chrysikos2015killingtwistorspinorstorsion} and 
references therein.}. If this is indeed the case, the spinors of this theory satisfy the \textit{zero-mode}/ \textit{Wilson line} type of solution:
\begin{align*}
i D_{\omega} \psi = 0 \quad \Rightarrow \quad \psi \left( x \right) = \mathbb{P} \exp \left( i \int_{\pi \left( x_0 , x \right)} \omega \right) \cdot \psi_0 
= \mathbb{P} \exp \left( i \int_{\pi \left( x_0 , x \right)} \bar{\omega} + \bar{J} \right) \cdot \psi \left( x_0 \right) \quad ,
\end{align*}
where $\mathbb{P}$ stands for the \textit{path ordered} operator, $\pi \left( x_0 , x \right)$ is a $1$-chain (a path) between the points $x_0$ and $x$ in  $\mathcal{M}$. If we take the path $\pi$ to be closed, i.e. $\pi \left( x_0 , x \right) \circ \pi' \left( x , x_0 \right) \cong \partial C$ (See Figure \ref{holonomyfig}), we recognize the operator acting on $\psi \left( x_0 \right)$ as the $x_0$ based holonomy:
\begin{align}\label{holcomp}
\text{Hol}_{x_0} \left( \omega \right) \cong \mathbb{P} \exp \left( i \int_{\partial C} \bar{\omega} + i \int_{\partial C} \theta \right) = \mathbb{P} \exp \left( i \int_{\partial C} \bar{\omega} \right) \cong \text{Hol}_{x_0} \left( \bar{\omega} \right) \quad \quad ,
\end{align}
where the last line is justified by homotopy invariance of the Lie algebra $\mathfrak{h}^*_{\theta}$ and by recalling that $\bar{J} \sim \theta$, the class of harmonic forms must be independent of the representation.

Consequently, the $x_0$ based holonomy for the Levi-Civita connection has the same information as that of the total connection $\omega$. Furthermore, since we are taking $\mathcal{M}$ to be simply-connected, we have that $\text{Hol}_{x_0} \left( \bar{\omega} \right) \cong \text{Hol}^0_{x_0} \left( \bar{\omega} \right)$, where the latter are the connected components of the holonomy based on $x_0$ and, therefore, we fall within the well-known classification for Lorentzian manifolds given by Leistner \cite{10.4310/jdg/1180135694}. In summary, this contortion does not alter the holonomy content of the theory, since the Ehresmann connection does not have spurious additional terms. Consequently, they will have the same loop representation if quantum effects are taken into account.

 \subsection{\label{subsec:killing} Killing equation and spinor effective mass.}


Let us study the Ehresmann parallel hypothesis in more depth. In light of the equation (\ref{torsion2}), we cast the parallel condition in the equivalent form: 
\begin{align}\label{parallelspinor2}
0 = i D_{\omega} \psi = i D_{\bar{\omega}} \psi_- + \left( i D_{\bar{\omega}} - \frac{i}{2} \iota_{J^{\sharp}} \left( \Sigma \right) \right) \cdot \psi_+ \quad \quad , 
\end{align} 
where we have defined the chiral spinors $\psi_{\pm} := P_{\pm} \cdot \psi$ as well as the operator $\Sigma := \sigma_{ab}\Sigma^{ab}$, we have used the identity $P_+ + P_- = \text{id}_S$ and the orthogonality relation $P_+ P_- = P_- P_+ =0$. Hence, from linear independence, the spin connection induces an explicit chiral symmetry breaking:
\begin{enumerate}
\item a chiral spinor $\psi_-$ satisfying a Levi-Civita-parallel field equation:
\begin{align}\label{wilsonlineLC}
\left. \begin{array}{rl}
i D_{\bar{\omega}} \psi_- &= 0 \\
\text{s.t} \;\; \left.\psi_- \right|_{x=x_0} &= \psi^0_-
\end{array}
\right\}
\quad \quad \Rightarrow \quad \quad 
\left\{ \begin{array}{rl}
& \psi_- \left( x \right) = \mathbb{P} \exp \left( i \int_{\pi \left( x_0 , x \right)} \bar{\omega} \right) \cdot \psi^0_- \\
& \text{satisfying} \quad \quad \square^{\bar{\omega}} \psi_- = 0
\end{array}\right.
\quad ,
\end{align}
which is also a \textit{Wilson line} type of solution\footnote{The existence of Levi-Civita-parallel spinors or, equivalently, torsion-less \textit{Ricci flat} spinors under the context of Lorentzian manifolds and Levi-Civita connections have been studied in several papers, such as \cite{Leistner2002,IKEMAKHEN2007299,Dikarev_2021,Ammann_2021} among others. We refer to those papers and the references therein for detailed discussions.}. Physically, this field represents a massless spinor with left chirality, and as such, it presents itself as a kind of \textit{Weyl fermion}. We have two reasons to discard the involvement of this field in this gravitational model. The first is that this field seems to be non-interacting and, as such, it becomes non-falsifiable as a physical field. The second will be apparent below.

\item a chiral spinor $\psi_+$ satisfying (the Killing-like) equation $i D_{\bar{\omega}} \psi_+ = \frac{i}{2} \iota_{J^{\sharp}} \left( \Sigma \right) \cdot \psi_+$ , with $\left. \psi_+ \right|_{x=x_0} = \psi^0_+$; which is usually written in its Dirac operator form:
\begin{align}\label{Diracopext}
i D^g \psi_+ = \frac{i}{2} \iota_{e^{\sharp}} \iota_{J^{\sharp}} \left( \Sigma \right) \cdot \psi_+ = - \frac{3}{2} J  \cdot  \psi_+ \cong - 3 s \, T \cdot \psi_+ \quad \quad .
\end{align}
This equation is consistent with 
the structure of parallel spinors with torsion for the generalized Dirac operator $D^s$ in equation (\ref{covdevger}), with $s = 1 / 2$ $\left( \frac{1}{2} J \cong  s T \right)$. Moreover, the existence of eigenvalues for the usual Dirac operator $ D\!\!\!\!/ \, = D^g$ in curved spaces is equivalent to the existence of eigenvalues for the torsion operator $T$ (specifically for the normalized parameter $s \to s/3$). This is, being $\mu$ the eigenvalue of the operator $T \cdot \psi_{\beta} = \mu \, \psi_{\beta}$, from (\ref{Diracopext}) with $s \to s/3$, the form:
\begin{align}\label{Diraceigen}
T \cdot \psi_+ = \mu \, \psi_+ \quad \Rightarrow \quad \left( i D\!\!\!\!/ \, + m_s \right) \cdot \psi_+ = 0 \quad \text{where} \quad \left. m_s \right|_{s = 1/2} = \frac{\mu}{2} \quad \quad ,
\end{align}
can be recognized as a Dirac-like equation with \textit{effective spinor mass $m_s$. This identification is only valid if $\mu \in \mathbb{R} \setminus \left\{ 0 \right\}$ (since $\mu = 0$ degenerates into the case \ref{wilsonlineLC}). In summary, the Killing equation \eqref{Diracopext} generates an effective mass when measured by a Levi-Civita connection.} The latter is somewhat reminiscent of the Higgs mechanism in the context of particle physics. 
\end{enumerate}

A word on the boundedness of equation (\ref{Diraceigen}) is needed. The latter involves the $\mu$ eigenvalue and, although the general case for its spectrum is not known (meaning that $\mu$ could eventually be a complex number) only bounded solutions for $\mu \in \mathbb{R} \setminus \left\{ 0 \right\}$ have been found \cite{chrysikos2015killingtwistorspinorstorsion}. We favor this case as a matter of informed guess\footnote{In fact, in the Riemannian case, $\mu$ is severely restricted to be real and constant, or a function with purely imaginary values \cite{Lichnerowicz1987}.}. In a more specific note, the plausibility of the latter discussion can be justified through the following technical results, where we will be using the notation of Appendix \ref{app:parallel}:
\begin{enumerate}
\item Since the restriction $\nabla^{s/3} T \cong \nabla^{s/3} J = 0$ is identically satisfied, we can split the spin bundle into eigenbundles with sections $\Sigma_{\mu}$ labeled by its eigenvalues $\mu$.

\item Our parallel spinor with torsion results in a null Ricci tensor ($\text{Ric} = 0$) and a null scalar curvature ($\text{Scal} = 0$) associated to the total Ehresmann connection  \cite[See Theorem 3.1][]{chrysikos2015killingtwistorspinorstorsion}. Thus, from equations (\ref{covdevger}) and (\ref{Lichnerowiczspinor}), the following are satisfied in $\Sigma_{\mu}$:
\begin{align}\label{eigenvalue1}
\Delta^{\bar{\omega}} + \frac{\bar{R}}{4} = s^2 \mu^2 \quad \quad , \quad \quad  \Delta^s + \frac{\bar{R}}{4} = 2 s^2 \left| J \right|^2_g \quad \quad  .
\end{align}
It follows from the latter, for $s \rightarrow s/3$, that:
\begin{align}\label{eigenvaluerestrictions}
\Delta^{1/2} - \Delta^{\bar{\omega}} = \frac{1}{2} \left( \left| J \right|^2_g - \frac{\mu^2}{2} \right) \quad \quad , 
\end{align}
which is an obstruction for the Levi-Civita Laplacian and the Ehresmann Laplacian to be equal and should be taken as a particular example of the Weitzenb\"{o}ck formula in this context. In order to still assume bounded solutions, for the eigenvalue expressions (\ref{eigenvalue1}) to be valid, we will restrict $\left| J \right|^2_g$ to be real.
\end{enumerate}

\subsection{\label{subsec:conserved} Lorentz Dirac current.}

Let us return to the Killing equation (\ref{Diraceigen}). Recall that the topological information in it is directly encoded through the explicit appearance of $\theta$ in $\bar{J}$. The parallel spinor hypothesis allows us to assume that $\psi = \psi_+$ (i.e. $\psi_- = 0$), motivated by the fact that $P_+ S$ is the only subsector of $S$ that explicitly carries the topological information coming from $[\theta]$. Thus, we can write  (\ref{Jcurrent}) in the form:
\begin{align}\label{Jalphacurrent}
\bar{J} \cong 
- C_{\alpha} \left\langle i P_+ e  \right\rangle_{\psi} = - C_{\alpha} \left\langle i e  \right\rangle_{\psi_+} \quad \quad \text{where} \quad \quad C_{\alpha} := \frac{2 \alpha i}{\alpha i + 1} \quad \quad ,
\end{align}
\Jav{which we also write in tensorial fashion for completeness:
\begin{align*}
\bar{J}_{\mu} = - C_{\alpha} \bar{\psi} i P_+ \gamma_{\mu} \psi = - C_{\alpha} \bar{\psi} \frac{i}{2} \left( 1 + \gamma_* \right) \gamma_{\mu} \psi \quad \quad ,
\end{align*}
where in the last equality 
we have used equation \eqref{chiralproj}.} In other words, the Barbero-Immirzi parameter $\alpha$ can be used to normalize the current $\bar{J}$ if needed. We justify such normalization arguing that $\bar{J}$ should be a Lorentzian Dirac current \cite[See][and the references therein]{IKEMAKHEN2007299,Dikarev_2021,Ammann_2021} so that physical meaning can be attached to it. The latter is characterized by $\eta \left( j , O \right) = - \left\langle O \right\rangle_{\psi}$, where $O \cdot \psi$ is a representation of $O \in T \left( \mathcal{M} \right)$ acting over $\psi \in S$. In our case $O \cong i \iota_{e^{\sharp}} \left( O \right) $, identified  as the generators of its associated Lie algebra. Additionally, $j$ must satisfy $\eta\left( j , j \right) = - \left| j \right|^2_g$ , with nullity only for $j = 0 $. Consequently, we can check directly that $\bar{J}$ in (\ref{Jalphacurrent}) satisfies:
\begin{align}\label{normalizedJcurrent}
\eta \left( \bar{J}^{\sharp} , O \right) 
= - \left\langle i \iota_{e^{\sharp}} \left( O \right) \right\rangle_{\psi_+} \quad \text{with} \quad \eta \left( \bar{J}^{\sharp}, \bar{J}^{\sharp}\right) 
= - \left| \bar{J} \right|^2_g \quad \text{for} \quad \alpha = -i \quad (C_{\alpha} = 1) \quad .
\end{align}
We remark that choosing $\alpha = -i$ is equivalent to the celebrated \textit{Ashtekar's choice} \cite{Vyas:2022etz} for the Barbero-Immirzi parameter. Notice that, prior to introducing the Ehresmann parallel spinor hypothesis in equation (\ref{Jcurrent}), this choice was forbidden. With such choice, from (\ref{normalizedJcurrent}) with $\alpha = -i$, we have $L_a \left( \bar{J} \right)  \sim d \iota_a J = d J_a = - D_{\omega} \left\langle i P_+ \gamma_a \right\rangle_{\psi} = 0$ by the parallel spinor hypothesis. This simplifies equation (\ref{upsilon}) to:
\begin{align}\label{upsilonfinal}
F_{ab} = i \star \left( \bar{J} \wedge K_{ab} \left( \bar{J} \right) \right)  = i \star \left( \bar{J} \wedge \iota_{\bar{J}^{\sharp}} \left( \mathcal{P}_+ \cdot \Sigma_{ab}^{\flat} \right) \right) = i B^{-2} \star \left( J \wedge \iota_{J^{\sharp}} \left( \mathcal{P}_+ \cdot \Sigma_{ab}^{\flat} \right) \right) 
\quad \quad ,
\end{align}
which will be used in the following section. Notice that this field strength $2$-form satisfies the anti-self-dual chiral symmetry $F^{(*)}_{ab} = -i F_{ab}$ and, consequently, the null property $F^2 \sim 0$, which would imply the existence of a Lorentzian chiral field (pure radiation).

\subsection{\label{summary:sec5} Summary of the section}

The spinor field decomposes into chiral components, through its conservation, the parallel condition, which leads to a modified Dirac equation (a Killing spinor equation) for one of its chiral component ($\psi_+$). This equation incorporates the topological current $J \sim \theta$ and gives the spinor an \textit{effective mass}, reminiscent of a Higgs-like mechanism. The other chiral component ($\psi_-$) remains massless and non-interacting, and is effectively discarded.

\Jav{The tensorial form of the simplified contortion field strength \eqref{upsilonfinal} can be written as:
\begin{align*}
F_{\mu \nu \alpha \beta} = - \frac{i}{4} g_{\zeta \mu} g_{\eta \nu} \bar{J}^{\delta} \left( \bar{J}^{\eta}  \epsilon^{\zeta}_{\;\; \delta \alpha \beta} - \bar{J}^{\zeta} \epsilon^{\eta}_{\;\; \delta \alpha \beta} \right) + \frac{1}{4} \epsilon^{\delta \gamma}_{\;\;\; \zeta \iota} g_{\gamma \mu} g_{\delta \nu} \bar{J}^{\eta} \left( \bar{J}^{\zeta} \epsilon^{\iota}_{\;\; \eta \alpha \beta} - \bar{J}^{\iota} \epsilon^{\zeta}_{\;\; \eta \alpha \beta} \right) \quad \quad .
\end{align*}}


\section{\label{sec:curvature} Curvature and generalized Einstein equations.}

In this section, we will address the remaining field equation, the one given by Eq. (\ref{fe11}). We aim to reconcile this with the discussion of the previous field equations and present a consistent picture of the gravitational model overall. We will argue that our result shows an effective theory that resembles GR with additional contributions that possess the expected qualitative behavior of \textit{Dark Energy} \Jav{\Mout{and some \textit{Dark fluid}}} associated with a topological origin. The mechanism of how this is achieved should be clear from the above sections; the key ingredient is the existence of an Ehresmann parallel spinor satisfying a Killing equation, as discussed in section \ref{sec:parallelspinors}. That spinor is ultimately responsible for measuring the topological degrees of freedom.

Let us begin by using the decomposition (\ref{curvaturedecomp}). It follows that we can cast Eq. (\ref{fe11}) into: 
\begin{align}\label{fe14}
\bar{R}^{(*)}_{ab} \wedge e^b + 2 \left( \frac{1}{2} \tilde{\alpha} \triangleright F_{ab} \right) \wedge e^b + \frac{1}{2} \left| J \right|^2_g \star e^{\flat}_a = - \frac{3}{2}  \left| d \phi \right|^2_g \star e^{\flat}_A \quad \quad,
\end{align}
where, given that we have set $\alpha = -i$ in the previous section, and using the $2$-form projectors of equation (\ref{chiralproj}), we write the operator $\frac{1}{2} \tilde{\alpha} \; \triangleright = -i \mathcal{P}_+$ . The symmetric part of the field strength $2$-form, shown in the first round bracket, can then be readily calculated to be:
\begin{align}\label{Fsymm}
\mathcal{P}_+ \cdot F_{ab} 
= i \frac{\left( \star - i \right)}{2} \frac{\left( \star + i \right)}{2} \cdot \left( \bar{J} \wedge i k^{\flat}_{ab} \left( \bar{J} \right) \right) + \frac{i}{2} \left| \bar{J} \right|^2_g \frac{\left( \star - i \right)}{2} \Sigma_{ab}^{\flat}
= \frac{i B^{-2}}{4}  \left| J \right|^2_g \Pi_{ab} \quad \quad,
\end{align}
where, in the last equality, we have inserted back the $B$ parameter by using again $\bar{J} = B^{-1} J$. When this is inserted back in (\ref{fe14}), after some algebra, we obtain the simplified equation:
\begin{align}\label{fe13}
\bar{R}^{(*)}_{ab} \wedge e^b + \frac{3}{2} \left( 1 + \frac{1}{B^2} \right) \left| J \right|^2_g \star e^{\flat}_a + L_{J^{\sharp}} \left( \phi \right) \star e^a = - \frac{1}{2} \left| d \phi \right|^2_g \star e^{\flat}_a \quad \quad \quad .
\end{align}
This is the version of the frame field equation that we will work with for the rest of the section.

\subsection{\label{subsec:orthogonality} Hodge decomposition Orthogonality.}

To further simplify equation (\ref{fe13}), we will make use of the orthogonality of the short Hodge decomposition (\ref{hodgedecomp}). Notice that in light of this, the pseudonorm of the Lorentz current is given by $\left| J \right|^2_g = \left| d \phi \right|^2_g +  \left| \theta \right|^2_g$ with $ \left\langle \theta , d \phi \right\rangle = L_{\theta^{\sharp}} \left( \phi \right) = \theta^{\sharp} \left( \phi \right) = \theta^{\mu} \partial_{\mu} \phi = 0$ . This suggests a potential of the form:
\begin{align}\label{orthogonality1}
\phi = \phi \left( u \right) \quad \text{where} \quad u = k \cdot x \quad \text{with} \quad k \in \mathbb{R}^{1,3} \quad \text{such that} \quad \theta^{\sharp} \cdot k = 0 \quad \quad .
\end{align}
Inserted back in the current pseudonorm we obtain:
\begin{align}\label{orthogonality2}
\left| J \right|^2_g = \phi'^2 \left| k^{\flat} \right|^2_g + \left| \theta \right|^2_g \quad \quad \text{where} \quad \quad  \phi' := d\phi / d u \quad \quad . 
\end{align}
As mentioned before, we will restrict to $\left| J \right|^2_g \in \mathbb{R}$ so that the expressions in (\ref{eigenvalue1}) can still be interpreted as restrictions on bounded spectra of differential operators. Moreover, if we combine the parallel spinor hypothesis with the potential of equation (\ref{orthogonality1}), we get:
\begin{align}\label{parellelspinorrest1}
\phi'' k^{\flat} \iota_a k^{\flat} + \phi' L_a \left( k^{\flat} \right) = - L_a \left( \theta \right) \;\; \text{which implies} \;\; \phi'' \left| k^{\flat} \right|^2_g + \phi' L^a \iota_a \left( k^{\flat} \right) = - L^a \iota_a \left( \theta \right) \;\; .
\end{align}
Using equations (\ref{fe4}), (\ref{Jalphacurrent}) with $\alpha = -i$ and  (\ref{torsion2}), we arrive at the expressions: 
\begin{align}\label{parellelspinorrest2}
d^* J 
= \frac{3}{2} \left| J \right|^2_g = \Delta \phi = \phi'' \left| k^{\flat} \right|^2_g + \phi' d^* k^{\flat} 
\quad \quad  \text{with} \quad \quad
d^* k^{\flat} 
= L^a \iota_a \left( k^{\flat} \right) + \frac{3i}{2} \phi' \left| k^{\flat} \right|^2_g \quad .
\end{align}
Combining now (\ref{parellelspinorrest1}) and (\ref{parellelspinorrest2}), we find the following general non-homogeneous Ricatti differential equation for the function $\phi' = d \phi / du$:
\begin{align*}
\phi'' \left| k^{\flat} \right|^2_g + \phi' L^a \iota_a \left( k^{\flat} \right) + \frac{3}{2} \left( i - 1 \right) \phi'^2 \left| k^{\flat} \right|^2_g - \frac{3}{2}\left| \theta \right|^2_g = 0 \quad \quad .
\end{align*}
However, since the solution of a Ricatti equation calls for the knowledge of a general global solution, we can turn around this difficulty by combining again  (\ref{parellelspinorrest1}) and (\ref{parellelspinorrest2}) with (\ref{orthogonality2}), to obtain: 
\begin{align*}
& \left| J \right|^2_g 
= \frac{ \left| \theta \right|^2_g - \frac{2}{3} L^a \iota_a \left( \theta \right)}{2} + \frac{\left| \theta \right|^2_g + \frac{2}{3} L^a \iota_a \left( \theta \right)}{2 i} \;\;\, \text{and} \,\;\; 
\phi'^2 \left| k^{\flat} \right|^2_g = - \frac{\left( 1 + i \right)}{2} \left( \left| \theta \right|^2_g + \frac{2}{3} L^a \iota_a \left( \theta \right) \right)  \;\;\; .
\end{align*}
It should be clear now that $\left| J \right|^2_g$ becomes a real quantity, considering $k$ as a null vector, or equivalently, if $\left| \theta \right|^2_g = - \frac{2}{3} L^a \iota_a \left( \theta \right)$. This restriction is met by topological considerations as follows: if we expand the harmonic form as $\theta = \theta_a e^b$, we can write $L^a \iota_a \left( \theta \right) = \eta^{ab} \nabla_a \theta_b$. It is then easy to see that the latter condition can also be obtained as a consequence of the fact that $\star \theta$ is also a harmonic form: $0 = d \star \theta = \star \left( \eta^{ab} \nabla_a \theta_b + \frac{3}{2} \left| \theta \right|^2_g \right)$. Summing up, the pseudonorm satisfies $\phi'^2 \left| k^{\flat} \right|^2_g = \left| d \phi \right|^2_g  = 0$, \Jav{that is, $d \phi^{\sharp}$ is light-like. Unlike more well-known abelian gauge theories like QCD, in this framework, a gauge symmetry breaking term like $\left| J \right|^2_g$ is not forbidden if the gauge field $\phi$ satisfies the light-like condition $\left| d \phi \right|^2_g  = 0$. This condition must be understood primarily as a gauge-fixing condition and secondly as a boundary condition that makes the field $\phi$ non-
propagati
ng.}

In consequence, $\left| J \right|^2_g \rightarrow \left| \theta \right|^2_g$ . Moreover, taking $k$ to be null allows for a non-trivial solution if and only if $\theta^{\sharp}$ is spacelike, i.e. $\left| \theta \right|^2_g > 0$ in our \textit{mostly negative} signature convention. Thus, inserting this back in (\ref{fe13}) yields:
\Jav{\begin{align}\label{fe15}
\left( \bar{R}^{(*)}_{ab} + \frac{3}{2} \left| \theta \right|^2_g \Sigma^{(*)}_{ab} \right) \wedge e^b = 0  \quad \quad , 
\end{align}}
where we have taken $B = 1$ for definiteness. This choice is ultimately unimportant since the harmonic $1$-form $\theta$ is defined up to homotopy and can always be renormalized. \Jav{Written in tensorial form we have:
\begin{align*}
\bar{R}_{\mu \nu} - \frac{1}{2} g_{\mu \nu} \bar{R}  + \left| \theta \right|^2_g g_{\mu \nu} = 0 \quad \quad ,    
\end{align*}}
\Jav{As the above form makes it appearent
, }equation (\ref{fe15}) is formally equivalent to GR with an effective \textit{Dark energy} term given by $\sim \left| \theta \right|^2_g > 0$, which is dynamical in nature. 
\Jav{This becomes more explicit when writing this term as $\left| \theta \right|^2_g = d^* \Lambda_{\xi}$, where we have used the expression \eqref{DNexpression}. Hence, the dark energy term carries information from the boundary of the space-time via $\xi$. We will comment on the specific topological origin of this term in the following subsection.}


\subsection{\label{subsec:mathystuff} Further comments on structure,  topology and bounded-ness}

From a mathematical point of view, we can understand the presence of the term $\left| \theta \right|^2_g$ in (\ref{fe15}) as the effect of the \textit{quadratic Casimir element} $C_2$ in the (co-)adjoint representation of the compact Lie algebra $\mathfrak{h}^*_{[\theta]}$\footnote{\vspace{-0.6cm}\begin{singlespace}    
To see this (quite informally), consider an infinitesimal interval $\left[ p , p + \delta p \right]$ so that the \textit{exponential diffeomorphism} $\exp_p : V \rightarrow U$, where $V$ is a neighborhood of $0$ in $T_p \mathcal{M}$ with $p \in \mathcal{M} $ and $U$ is a neighborhood of $p$ in $\mathcal{M}$. This map acts as $\Phi_1^{i \iota_X \left( \theta \right)} \left( 1 \right) \simeq 1 + i \iota_{X \left( p \right)} \left( \theta \right) \delta p $, where $\Phi_1^{i \iota_X \left( \theta \right)}$ is the flow generated by the generators $i \iota_X \left( \theta \right)$ in the direction of the isotropy group $H_{[\theta]}$ and is known to be a surjective map.
\end{singlespace}}. The latter must be understood as coming from an \textit{abelian current algebra} associated to the chiral current $J$ of equation (\ref{Jalphacurrent}), which is known to encode critical physical data like scaling dimensions and anomalies in the context of QFT. Moreover, since the field strength (\ref{upsilonfinal}) is anti-self-dual, it has the null property $F^2 \sim 0$. This means, by the APS theorem on manifolds with torsion \cite[See, for instance,][and the references therein]{Mason:1991rf,Bleecker:2013bka,Kobayashi_2021} that the index of the Dirac operator $i D$ has a non-trivial contribution from the Nieh-Yan term (\ref{NiehYan}). In other words, the chiral anomaly can be related exclusively to the latter. The complete calculation is highly non-trivial and should be analyzed in a case by case for solutions of the frame equation (\ref{fe15}) above. Nevertheless, from (\ref{torsion2}) we can directly calculate the Nieh-Yan contribution to be:
\begin{align}\label{NYcalc}
\int_{\mathcal{M}} C_N 
\sim \int_{\partial \mathcal{M}} j^*_{\partial} \left(  K_{ab} \wedge \Sigma^{ba} \right) 
\sim \int_{\mathcal{M}} d \star J \sim \int_{\mathcal{M}} \star \left| \theta \right|^2_g \quad \quad ,
\end{align}
which confirms our suspicions. \Jav{In addition, 
from (\ref{NiehYan}), when combining this result with \eqref{DNexpression}, we can see that $K_{ab} \wedge \Sigma^{ba}$ is homotopic to the Dirichlet to Neumann operator bilinear $3$-form $\xi \Lambda \xi$, which has been discussed to be topological in subsection \ref{subsec:BVproblem}.}

Notice that this quantity is completely topological and bounded from below, since it is coming from a normed Lorentzian Dirac current. On the same note, the square of the Dirac operator ($D^2 \sim \Delta^{\bar{\omega}}$) is generally bounded from below\footnote{This is a direct consequence of the special geometry enforced by the parallel spinor condition (e.g., vanishing scalar curvature in pp-waves).}. Hence, from (\ref{eigenvaluerestrictions}), we have:
\begin{align}\label{mineigenvalue}
\mathbb{R} \ni C := \min_{x \in \mathcal{M}} \left( \Delta^{\bar{\omega}} + \frac{1}{2} \left| \theta \right|^2_g \right) = \min_{x \in \mathcal{M}} \left( \Delta^{1/2} + \left( \frac{\mu}{2} \right)^2 \right) \quad \quad .
\end{align}

\subsection{\label{summary:sec6} Summary of the section}

Inserting all the previous elements into the frame field equations, the topological terms contribute as effective energy-momentum components. The final simplified gravitational equation includes a positive, dynamical \textit{Dark Energy} component (a cosmological constant-like term, but originating from topology)\Jav{\Mout{and a stress-energy tensor associated with a scalar field, interpreted as a \textit{Dark Fluid} component}}. \Jav{The field Eqs. in the GR driven regions then take the form of Eq.~\eqref{fe15}, which takes the following tensorial form: 
\begin{align*}
\bar{R}_{\mu \nu} - \frac{1}{2} g_{\mu \nu} \bar{R}  + \left| \theta \right|^2_g g_{\mu \nu} = 0 \quad \quad ,    
\end{align*}
and which makes the physical interpretation more direct.}

\section{\label{sec:final} Final Comments}

In this paper, we have introduced a mechanism of topological origin that could account for the presence of (dynamical) Dark Energy \Jav{\Mout{and some Dark Fluid}}. The framework presented attempts to unify the description of gravity (including torsion), matter (spinors), and the dark sector under a single topological mechanism. The model makes a concrete prediction: the dark energy density is not a true constant but is dynamic, tied to the harmonic $1$-form $\theta$. Its value could, in principle, be calculated from the number and properties of the topological defects (black holes) in the universe. This provides a potential pathway to falsification.

The most interesting result of this paper is that a major component of the universe's energy content, dark energy, is proposed not as a fundamental field or a constant, but as a manifestation of the universe's global topology\Jav{, as it is homotopic to the Nieh-Yan topological 0-form}. It turns out to be dynamical and bounded, which seems to be in accordance with recent observations. \Jav{Moreover, we find that it is not only related to the Nieh-Yan form, but also relates to 
the boundary conditions of the manifold for the harmonic one form $\theta$ via the Dirichlet to Neumann operator, which connects 
both kinds of boundary conditions. This is, to the best of our knowledge, the first time this operator gives a physical effect in gravitation. As such, this is reminiscent of a holographic effect, in the sense that the information of the dark energy term is stored at the boundary.\Mout{The appearance of a Dark Fluid requires further analysis}} The next steps involve exploring the cosmological implications of this model, solving the field equations in a cosmological context, and understanding the observational signatures of this topologically driven theory.

In conclusion, this paper offers a compelling mechanism that derives dark energy from the interplay between spacetime topology and quantum spinor fields, without introducing new fundamental entities.



\appendix

\section{\label{app:form} Tetrad formalism, differential forms and Cohomology with boundary}

Let $\Omega^k \left( \mathcal{M} \right)$ be the space of smooth complex exterior differential forms of degree $k$ over a $4$-manifold $\mathcal{M}$ and $\Omega^k \left( \mathcal{M} \right)^* $  its dual. Let $\Omega \left( \mathcal{M} \right) = \oplus^4_{k=0} \Omega^k \left( \mathcal{M} \right) $ its graded algebra. The usual operators on $\Omega \left( \mathcal{M} \right)$ are well defined: 
\begin{enumerate}
    \item The exterior derivative $d: \Omega^k \left( \mathcal{M} \right) \rightarrow \Omega^{k+1} \left( \mathcal{M} \right)$ with $0 \leq k \leq 3$,
    \item The exterior co-derivative $d^{\star}: \Omega^k \left( \mathcal{M} \right) \rightarrow \Omega^{k-1} \left( \mathcal{M} \right) $ with $1 \leq k \leq 4$ and
    \item The Hodge dual $\star: \Omega^k \left( \mathcal{M} \right) \rightarrow \Omega^{4-k} \left( \mathcal{M} \right)$ with $0 \leq k \leq 4$.
\end{enumerate}

Since we have assumed that our space-time manifold $\mathcal{M}$ is compact but not closed, i.e. with a non-trivial boundary $\partial \mathcal{M} \neq \emptyset$. We begin by recalling that, given the Hodge dual $\star: \Omega^k \left( \mathcal{M} \right) \rightarrow \Omega^{n-k} \left( \mathcal{M} \right)$, the operations ($s = -1 $)
\begin{align}\label{hodge}
\star \star = \left( - 1 \right)^{k\,(n-k)} s \;\;\; , \;\;\; \star d^{\star} = \left( - 1 \right)^k s d \star \;\; \; , \;\;\; \star d = \left( - 1 \right)^{k+1} s d^{\star} \;\;\;  \text{on} \;\;\; \Omega^k \left( \mathcal{M} \right) \;\;,
\end{align}
are well defined. We have that, in any Riemannian manifolds, the $L^2$-inner product, dependent on the metric $g \left( \cdot  , \cdot \right)$ induced by the Hodge dual for all $\alpha , \beta \in \Omega^k \left( \mathcal{M} \right)$ is naturally defined in the following way:
\begin{align}
\label{L2inner} \left( \alpha , \beta \right) = \int\limits_{\mathcal{M}} \alpha \wedge \star \beta = \int\limits_{\mathcal{M}} \left\langle \alpha , \beta \right\rangle d \mu 
\quad \text{and the $L^2$-type norm} \quad 
\left\| \alpha \right\|^2_g = \left( \alpha , \alpha \right) \quad \quad ,
\end{align}
are well defined. In our case we will use the following $p$-forms in a recurrent way:
\begin{table}[h!]
\begin{centering}
\begin{tabular}{|c|c|c|c|}
\hline 
$e^{a} = e^a_{\mu} dx^{\mu} 
$  & vierbein basis in $T^*\left(M\right)$ & $ \;\; \eta_{ab}:=g\left(e_{a},e_{b}\right) 
\;\; $  & Minkowski metric \tabularnewline 
$e_{a} = e_a^{\mu} \partial_{\mu} 
$  & vector basis in $T\left(M\right)$ & $\Sigma^{ab}:=\frac{1}{2}e^{a}\wedge{e}^{b}
$  & Plebanski $2$-form \tabularnewline
$\; d\mu:=\frac{1}{3} \Sigma^{ab} \wedge\Sigma_{ab}^{(*)}$  \;& $\star\left(1\right):=d\mu = \sqrt{-g} d^4x $ & $\bar{\omega}_{\;\;b}^{a} \in\Omega^{1}\left(M\right)
$  & Levi-Civita connection \tabularnewline
\hline 
\end{tabular}
\par\end{centering}
\caption{\label{tab:VEP-geometrical-elements} Geometrical data}
\end{table}

Since our case is Lorentzian, we thus take the $L^2$-inner-product-like expression as being merely scalar products $\left\langle \cdot , \cdot \right\rangle : \Omega^1 \left( \mathcal{M} \right) \times \Omega^1 \left( \mathcal{M} \right) \rightarrow \mathbb{C}$ over the tetrad with a \textit{mostly negative} signature convention $\left(+---\right)$. In other words, locally, i.e. in Minkowski space we have $\eta^{ab} := g \left( e^a , e^b  \right) = \left\langle e^a , e^b \right\rangle$ for the basis, taken to be real and can be naturally extended to $n$-forms by linearity. 
The latter allows for the definition of the musical isomorphisms between differential forms and tangent vectors $\flat: \Omega^k \left( \mathcal{M} \right) \rightarrow \Omega^k \left( \mathcal{M} \right)^*$ given by $ X^{\flat} \left( Y \right) = \left\langle X , Y \right\rangle $ and $\sharp: \Omega^k\left( \mathcal{M} \right)^* \rightarrow \Omega^k \left( \mathcal{M} \right)$ given by $\left\langle \omega^{\sharp} , Y \right\rangle = \omega \left( Y \right) $.

In the case where $\mathcal{M}$ has no boundary, it is easy to show that the exterior derivative $d$ and the exterior co-derivative $d^*$, operators defined above, are dual with respect to the inner product (\ref{L2inner}). This is no longer the case on manifolds with boundary. Let $\alpha \in \Omega^{k-1} \left( \mathcal{M} \right)$ and $\beta \in \Omega^{k} \left( \mathcal{M} \right)$ for $k \geq 1$. We then have:
\begin{align}\label{dualboundary}
\int_{\partial \mathcal{M}} j^*_{\partial} \left( \alpha \wedge \star \beta \right) = \left( d \alpha , \beta \right) - \left( \alpha, d^* \beta \right) \quad \quad .
\end{align}
where $j^*_{\partial}: \Omega^k \left( \mathcal{M} \right) \rightarrow \Omega^k \left( \partial \mathcal{M} \right)$ is the \textit{inclusion in the boundary operator}. However, duality can be recovered for particular boundary conditions over the differential forms, such as $j^*_{\partial} \left( \alpha \right) = 0$ or $j^*_{\partial} \left( \star \beta \right) = 0$ . When performing the variation of these differential structures naturally, well-posedness leads to considering the following metric compatibility conditions which will use extensively: $\eta^{ab} \eta_{ab} = \delta^a_{\;\;b} = \iota_b \left( e^a \right)$ and $d_{\omega} \left( \eta_{ab} \right) = 0$ . This allows for the rising and lowering of spin indices. Under the same context, the Lie derivative over differential forms is defined as usual: 
\begin{align}\label{Liederiv}
L_{a} \left( \omega \right) := \iota_{a} d \omega + d \iota_{a} \omega \quad \quad , \quad \quad L^{a}  \left( \omega \right) := \eta^{ab} L_b \left( \omega \right) \quad \quad \omega \in \Omega \left( \mathcal{M} \right) \quad \quad ,
\end{align}
while the second expression is induced by the metric. This primary picture is complemented with the first and second Bianchi's identities, respectively:
\begin{align}
\label{Bianchi} d_{\omega} T^a = R^a_{\;\;b} \wedge e^b \quad \quad , \quad \quad d_{\omega} R^a_{\;\;b} = 0 \quad \quad ,
\end{align}
which hold for any $T^a$ torsion $2$-form and $R^a_{\;\;b}$ curvature $2$-form defined from the same connection $1$-form $\omega$. Finally, on an oriented space-time manifold $\mathcal{M}$ taken as a compact $4$-manifold, we can define Chern type class called the \text{Nieh-Yan}: 
\begin{align}\label{NiehYan} 
C_N := d \left( e^a \wedge T_a \right) = T^a \wedge T_a - 2 R_{ab} \wedge \Sigma^{ab} \quad  \in H^{4} \left( \mathcal{M} ; \mathbb{Z} \right) \quad \quad , 
\end{align}
which is consistent with the appearance of torsional degrees of freedom \cite{Kaul,Babourova,Zanelli,Liko}.

Any smooth differential $p$-form has a natural decomposition into
tangential and normal components along the boundary of $\partial \mathcal{M}$ which, we can write $\omega \left( x \right) = \omega_{\text{tan}} \left( x \right) + \omega_{\text{norm}} \left( x \right)$ with $x \in \partial \mathcal{M}$. We write as $\Omega^p_N$ the space of smooth $p$-forms on $\mathcal{M}$ satisfying Neumann boundary conditions at every point of $\partial \mathcal{M}$ and, let $\Omega^p_D$ be the space of smooth $p$-forms on $\mathcal{M}$ satisfying Dirichlet boundary conditions at every point of $\partial \mathcal{M}$. This is:
\begin{align}\label{Neumann&Dirichlet}
\Omega^p_N = \left\{ \omega \in \Omega^p \; 
| \; \omega_{\text{norm}} = 0\right\} \quad \text{and} \quad \Omega^p_D = \left\{ \omega \in \Omega^p \;
| \; \omega_{\text{tan}} = 0 \right\} \quad \text{, respectively} \quad.
\end{align}
Thus defined, we write $cE^p_N = d^* \left( \Omega^{p+1}_N \right)$ and $E^p_D = d \left( \Omega^{p-1}_D \right)$, such that the boundary conditions are taken before the co-differential and differential operator. Going further; consider $\mathcal{M}$ an orientable compact manifold of dimension $n$ with boundary $ \partial \mathcal{M}$ ( as it is our case). Let $z \in H_{n} \left( \mathcal{M} ,\partial \mathcal{M} ; \mathbb {Z} \right)$ be the fundamental class of the manifold $\mathcal{M}$, then the cap product with $z$ (or its dual class in cohomology) induces a pairing of the (co)homology groups of $\mathcal{M}$ and the relative (co)homology of the pair $(M,\partial M)$. This gives rise to isomorphisms of $ H^{k} \left( \mathcal{M} ,\partial \mathcal{M} ; \mathbb {Z} \right)$ with $ H_{n-k}(M;\mathbb {Z} )$ and $H_{k} \left( \mathcal{M} ,\partial \mathcal{M} ; \mathbb {Z} \right)$ with $H^{n-k} \left( \mathcal{M} ; \mathbb {Z} \right)$ for all $\partial \mathcal{M}$, so Poincaré duality appears as a special case of the Lefschetz duality. For $A$ and $B$ subspaces of $\mathcal{M}$ with common boundary, for each $k$, there is an isomorphism $H^{k} \left( \mathcal{M}, A ; \mathbb {Z} \right) \to H_{n-k} \left( \mathcal{M} , B ; \mathbb {Z} \right)$.

\section{\label{app:parallel}Parallel spinors and torsion 
}

Let us define a covariant derivative associated to the $1$-form $J \in [\theta]$ as:
\begin{align}\label{thetacovdev}
\nabla^{J}_X \psi = \nabla^g_X \psi +  \iota_X \left( J \right) \cdot \psi \quad \text{with its h.c. as the corresponding  expression for $\bar{\psi}$} \quad , 
\end{align}
where $\nabla^g_X$ is the Levi-Civita connection, while $\iota_X \left( J \right) : S \rightarrow S$ are generators of the Lie group $h^*_{[\theta]}$ acting over the spinor field (in some given representation). Moreover, if the spinor is \textit{Levi-Civita-parallel}, i.e. parallel transported by the Levi-Civita connection along any path, manifested by $\nabla^g_X \psi = 0$, the $J$-covariant derivative is but the direct $\mathfrak{h}^*_{[\theta]}$-action over the spinors and carries along topological information. Now, as shown in  \cite{agricola2013twistorialeigenvalueestimatesgeneralized,chrysikos2015killingtwistorspinorstorsion} among others, these expressions can be thought of as being a consequence of a generalized torsion $T$:
\begin{align}\label{covdevger}
\nabla^s_X \psi := \nabla^g_X \psi + s \, \iota_X \left( T \right) \cdot \psi \quad \quad \text{with Dirac op.} \quad \quad D^s \psi = D^g \psi + 3 s T \cdot \psi \quad \quad .
\end{align}
The last expression gives a Schr\"odinger-Lichnerowicz formula of the type \cite{Bismut1989}:
\begin{align}\label{Lichnerowiczspinor}
\left( i D^{s/3} \right)^2 \cong \square^{s/3} \cong \Delta^{s/3} + s dT + \frac{1}{4} \bar{R} - 2s^2 \left| T \right|^2_g \quad \quad ,
\end{align}
where $\square^{s/3} \sim \square^{\omega} := \eta^{ab} \nabla^{\omega}_a \nabla^{\omega}_b$ is the D'Alembertian operator associated to the connection $\omega$, with $\nabla^{\omega}_a \alpha := \iota_a D_{\omega} \alpha$, while $\Delta^s = {\nabla^{s/3}}^* \nabla^{s/3}$ and $\bar{R}$ is the Ricci scalar curvature associated to the Levi-Civita connection. 
For completeness, equation (\ref{Lichnerowiczspinor}) latter should be contrasted with the usual Lichnerowicz-Weitzenb\"{o}ck formula  \cite{griffiths2014principles}: $\left( D^{s/3} \right)^2 = \Delta^{s/3} + \frac {1}{4} R + \frac {1}{2} F_{\omega}^{+}$ , where $F^{+}$ is the self-dual part of the field strength $2$-form acting via Clifford multiplication. 
The latter can be directly compared with (\ref{Lichnerowiczspinor}) when $s=1/2$. 

\section*{Acknowledgements}

J. L. E. thanks the FONDECYT grant N° 11241170 for the support during this research. Y. V. acknowledges the support of FONDECYT grant N° 1220871 throughout this work. MLeD acknowledges the financial support by the Lanzhou University starting fund, the Fundamental Research Funds for the Central Universities (Grants No. lzujbky-2019-25 and lzujbky-2025-jdzx07), the Natural Science Foundation of Gansu Province (No. 22JR5RA389 and No.25JRRA799), National Science Foundation of China  (NSFC grant No.12247101)
and the ‘111 Center’ under Grant No. B20063.

We dedicate these pages to our late colleague, Dr. Francisco Pe\~{n}a. 


\bibliography{sample}






\end{document}